\documentclass[
superscriptaddress,
reprint,
showpacs,preprintnumbers,
amsmath,amssymb,
aps,
prb,
floatfix,
]{revtex4-1}

\usepackage{graphicx,epsfig}
\usepackage{dcolumn}
\usepackage{bm}

\usepackage{mwe}

\usepackage{hyperref}
\hypersetup{colorlinks=true, linkcolor=blue, citecolor=blue, urlcolor=blue}

\begin{document}

\title{Magnetic and structural properties of the iron silicide superconductor LaFeSiH}

\author{M. F. Hansen} 
\affiliation{CNRS, Universit\'e Grenoble Alpes, Institut N\'eel, 38042 Grenoble, France}

\author{S. Layek} 
\affiliation{CNRS, Universit\'e Grenoble Alpes, Institut N\'eel, 38042 Grenoble, France}
\affiliation{Department of Physics, School of Advanced Engineering, University of Petroleum and Energy Studies, Dehradun, Uttarakhand 248007, India}

\author{J.-B. Vaney} 
\affiliation{Univ. Bordeaux, CNRS, Bordeaux INP, ICMCB, UMR 5026, F-33600 Pessac, France}

\author{L. Chaix}
\affiliation{CNRS, Universit\'e Grenoble Alpes, Institut N\'eel, 38042 Grenoble, France}

\author{M. R. Suchomel}
\affiliation{Univ. Bordeaux, CNRS, Bordeaux INP, ICMCB, UMR 5026, F-33600 Pessac, France}

\author{M. Mikolasek}
\author{G. Garbarino}
\author{A. Chumakov}
\author{R. R\"uffer}
\affiliation{ESRF-The European Synchrotron, CS40220 38043 Grenoble Cedex 9 France}

\author{V. Nassif}
\affiliation{CNRS, Universit\'e Grenoble Alpes, Institut N\'eel, 38042 Grenoble, France}
\affiliation{Institut Laue -Langevin, 71 Avenue des Martyrs, 38000 Grenoble, cedex 9 France}

\author{T. Hansen}
\affiliation{Institut Laue -Langevin, 71 Avenue des Martyrs, 38000 Grenoble, cedex 9 France}

\author{E. Elkaim}
\affiliation{Synchrotron SOLEIL, L'Orme des Merisiers Saint-Aubin, France}

\author{T. Pelletier}
\affiliation{CNRS, LNCMI, Univ. Grenoble Alpes, INSA-T, UPS, EMFL, Grenoble, France}

\author{H. Mayaffre}
\affiliation{CNRS, LNCMI, Univ. Grenoble Alpes, INSA-T, UPS, EMFL, Grenoble, France}

\author{F. Bernardini}
\affiliation{Dipartimento di Fisica, Universit\`a di Cagliari, Cittadella Universitaria, I-09042 Monserrato (CA), Italy}

\author{A. Sulpice}
\author{ M. N\'u\~nez-Regueiro} 
\author{P. Rodi\`ere} 
\author{A. Cano} 
\affiliation{CNRS, Universit\'e Grenoble Alpes, Institut N\'eel, 38042 Grenoble, France}

\author{S. Tenc\'e} 
\affiliation{Univ. Bordeaux, CNRS, Bordeaux INP, ICMCB, UMR 5026, F-33600 Pessac, France}

\author{P. Toulemonde}  
\affiliation{CNRS, Universit\'e Grenoble Alpes, Institut N\'eel, 38042 Grenoble, France}

\author{M.-H. Julien}
\affiliation{CNRS, LNCMI, Univ. Grenoble Alpes, INSA-T, UPS, EMFL, Grenoble, France}

\author{M. d'Astuto}
\affiliation{CNRS, Universit\'e Grenoble Alpes, Institut N\'eel, 38042 Grenoble, France}

\date{\today}

\begin{abstract}
The magnetic and structural  properties of the recently discovered pnictogen/chalcogen-free superconductor LaFeSiH ($T_c\simeq10$~K) have been investigated by $^{57}$Fe  synchrotron M{\"o}ssbauer source (SMS) spectroscopy, x-ray and neutron powder diffraction and $^{29}$Si nuclear magnetic resonance spectroscopy (NMR). No sign of long range magnetic order or local moments has been detected in any of the measurements and LaFeSiH remains tetragonal down to 2 K. The activated temperature dependence of both the NMR Knight shift and the relaxation rate $1/T_1$ is analogous to that observed in strongly overdoped Fe-based superconductors. These results, together with the temperature-independent NMR linewidth, show that LaFeSiH is an homogeneous metal, far from any magnetic or nematic instability, and with similar Fermi surface properties as strongly overdoped iron pnictides. This raises the prospect of enhancing the $T_c$ of LaFeSiH by reducing its carrier concentration through appropriate chemical substitutions. Additional SMS spectroscopy measurements under hydrostatic pressure up to 18.8~GPa found no measurable hyperfine field.

\end{abstract}


\maketitle

\section{Introduction} 
Following the discovery of Fe-based superconductors in 2008, the results of intense research effort has unearthed many compounds, leading to what now forms a complex and rich family of unconventional superconducting materials \cite{hosono-15,MARTINELLI20165,hosono-20}. 
In these systems, the superconducting instability is believed to be intimately linked to antiferromagnetic (AFM) fluctuations. 

Here, we focus on the compound LaFeSiH, for which superconductivity has been revealed, with $T_c=10$~K \cite{bernardini-lafesih-prb1}, i.e. at a value rather similar to other superconducting Fe-silicides discovered more recently, LaFeSiF$_{x}$\cite{LaFeSiF} and LaFeSiO$_{1-\delta}$ \cite{Hansen2022}. LaFeSiH was reported to undergo an orthorhombic distortion at moderate pressures and low temperatures \cite{bernardini-lafesih-prb1}.
According to first-principles calculations, superconductivity in LaFeSiH is unconventional in the sense that $T_c$ cannot be explained by the conventional electron-phonon pairing mechanism \cite{yildirim18-prb}. 
Furthermore, the measured magnetic penetration depth reveals the emergence of a $d$-wave superconducting gap in this system \cite{bhattacharyya2019}. 
This circumstance can be related to the increased orbital overlap along the $c$-axis together with the change in the orbital content of the Fermi surface compared to the reference LaFeAsO, which otherwise should favor the $s_\pm$-wave state. 
At the same time, despite the reduced Fermi-surface nesting, first-principles calculations suggest that the tendency towards magnetism in LaFeSiH is quite similar to LaFeAsO \cite{bernardini-lafesih-prb1,yildirim18-prb,arribi2020} as well as the loss of superconductivity at moderate pressures, associated with the decrease of the magnetic moment on iron, with an additional re-entrant behaviour as pressure is increased.  

In this paper we investigate the magnetic properties of LaFeSiH as a function of temperature, magnetic field, and pressure using SMS ($^{57}$Fe Synchrotron M{\"o}ssbauer Source), NPD (neutron powder diffraction), and NMR (nuclear magnetic resonance spectroscopy), as well as the low temperature structural properties using XRD (X-ray diffraction).

Our measurements show no long range magnetic ordering or local moments within detection limits Indeed, comparison of our present results using SMS in LaFeSiH with $^{57}$Fe M\"ossbauer literature for Fe-based superconductors\cite{KlaussPRL.101.077005LaFeAsO,McGuirePRB.78.094517LaFeAsO,KitaoJPSJ2008,Mizuguchi2010} suggests a upper bound of 0.04~$\mu_B$ for an ordered magnetic moment, in agreement with NMR which, under similar hypothesis, suggest an upper bound of $\sim$ 0.01~$\mu_B$.
This is well below the value of (DFT) first-principles calculations (1.2 $\mu_B$) \cite{bernardini-lafesih-prb1}.
Furthermore, SMS spectroscopy do not find any re-entrant magnetism at higher pressure up to 18.8 GPa. 
Based on the temperature dependence of 1/T$_{1}$, and consistent with the absence of magnetism, LaFeSiH seems located in the overdoped regime.
 
\section{Methods}

\subsection{Samples}

LaFeSi precursor powders were first prepared by arc melting of a stoichiometric mixture of pure elements (La, Fe and Si), subsequently grinded, compacted, and subjected to a thermal treatment of 7 days at 950~$^{\circ}$C. The final LaFeSiH powders were obtained from the hydrogenation of the precursors, treated at 250~$^{\circ}$C for 4~h under a static pressure of gaseous H$_2$ of 10 bars. The deuterated sample, LaFeSiD, for the NPD experiment was obtained with a similar protocol, hydrogen gas being replaced by deuterium gas in the last stage (see also \cite{HANSEN2023169281}). For the preparation of LaFeSiH single crystals, small single crystals of LaFeSi were separated from the bulk of an arc-melted Si-under-stoichiometric LaFeSi$_{0.86}$ composition. The 20 $\mu$m thick plate-like single crystals of various sizes were then hydrogenated at 250~$^{\circ}$C under a flow of H$_2$ gas for 4~h. Both the LaFeSiH powders and single crystals were checked by X-ray diffraction and refinement of the unit cell parameters were found to be in good agreement with previously reported data \cite{bernardini-lafesih-prb1}. For the SMS measurements, additional samples were synthesized using 96\% isotopic Fe$^{57}$ instead of the naturally abundant isotope.
Bulk superconductivity was confirmed by measuring the magnetization of all samples, as described in Appendix C.  
Furthermore, the superconducting state was verified using resistivity measurements for the isotope enriched sample, as described in Appendix B.

\subsection{SMS spectroscopy under pressure and magnetic field}\label{Mossmeth} 

Standard $^{57}$Fe M\"ossbauer spectroscopy with laboratory sources has been extensively applied to investigate structural, electronic and magnetic properties of iron-based pnictides and chalcogenides \cite{RevMossbauerPnictides}.
In order to probe the magnetic phase diagram of LaFeSiH at both ambient and high pressure, we instead choose SMS spectroscopy using a synchrotron source to allow the combination of both low temperature, high pressure and high magnetic field. 
While combining high-pressure with low temperature and high magnetic field bears no novelty using standard M\"ossbauer spectroscopy with laboratory sources \cite{FlutzRevMoss} or Nuclear Forward Scattering \cite{LubbersHI1999}, the possibility to use SMS spectroscopy at these conditions has emerged only recently, with relatively few publications relating this option \cite{PhysRevB.99.134443,PhysRevB.101.094433, PhysRevB.105.054417}. This despite the advantage of having a very focused beam, already in use since its beginning for standard high pressure experiments \cite{Potapkin:vv5038}. 
We used the $^{57}$Fe Synchrotron M\"ossbauer Source (SMS) \cite{PhysRevB.55.5811,Potapkin:vv5038} at the Nuclear Resonance beamline (ID18) \cite{ID18ref} of the European Synchrotron Radiation Facility (ESRF, Grenoble), where this was developed. The $^{57}$FeBO$_3$ crystal was set in the (111) reflection and sinusoidal acceleration configuration as described in Ref. \onlinecite{Potapkin:vv5038}. 
Energy calibrations were performed for each velocity configuration of the Doppler spectrometer, using a 25 $\mu$m thick natural $\alpha$-iron absorber standard. We used a maximum velocity of 11.30(2) mm/s for the first set of measurements, at low field and ambient pressure, and 5.64(1) mm/s for the remaining measurements. Here we detail the measurements of two samples, one with isotope enriched iron and one with natural iron. This is further detailed in Appendix \ref{SMSadd}.

Data analysis was carried out with the MossA procedure for Matlab\copyright \cite{Prescher:he5543}. 

An example of the data obtained in LaFeSiH at low temperature is shown in Fig. \ref{Mossb_LT_ambP}, at ambient pressure and without any applied field, as further detailed in Sec. \ref{MossRes} (raw data are shown in Fig. \ref{Mossb_LT_ambP-raw}, see Appendix \ref{SMSadd}).  

For all measurements the cryomagnet system of the beamline was used, providing temperatures between 2 and 300 K and an external magnetic field between -8 and +8 T in vertical direction. The system is equipped with aluminum windows for the $\gamma$-ray beam and quartz windows for the ruby fluorescence calibration signal. Pressure calibration was performed in-situ after each change of pressure and/or temperature conditions.

A membrane driven diamond anvil cell (DAC) with 500 $\mathrm{\mu}$m culet diamonds and stainless steel gasket was used.
 The enriched and natural sample were loaded into the same DAC, used for the temperature and field measurements at ambient pressure (as it can be seen in Fig. \ref{LaFeSiHinMDAC} in Appendix \ref{SMSadd}), using He as a pressure-transmitting medium. 
Using an optics system with a very long working distance (f $\frac{1}{28}$ 50 cm) we measured in-situ the ruby fluorescence for pressure calibration.

\subsection{Diffraction techniques}

Neutron powder diffraction (NPD) was carried out at the D2B powder diffraction instrument of the Institute Laue Langevin (ILL) using a wavelength of $\lambda$ = 1.5944 \AA{}. The deuterated sample, LaFeSiD, was inserted in a cylindrical vanadium sample holder and measured at 10 fixed temperature points in the 10~K - 300 K range with an exposure time of $\sim$ 2h per point and $\sim$ 4 h at 2 K. The empty cryostat was also measured, showing only one significant background peak at 151.2~deg.
X-Ray powder diffraction (XRD) was carried out at the CRISTAL beamline of the synchrotron SOLEIL. The sample was loaded in a 300 $\mu$m diameter borosilicate capillary. Two different detectors were used in parallel for each: a fast photon counting strip detector (Mythen2) for medium angular resolution but excellent counting statistics acquisition and a multi-crystal analyser for high angular resolution and medium counting statistics acquisition. Precise lattice parameters were determined using the multi-crystal analyser data and the accurate atomic positions were obtained from refinements of the Mythen2 data. The wavelength used for the experiment was $\lambda$ = 0.51302 \AA{}. The collected data was corrected for absorption using a $\mu$R value of 1.23.
XRD and NDP results are compared in appendix D.

\subsection{Nuclear magnetic resonance}
$^{29}$Si NMR experiments were performed in an external field of $\sim$15~T (the exact field value is calibrated from the $^{63}$Cu NMR resonance frequency in metallic Cu) down to 2.85~K. At this field, the superconducting transition is not detected in our measurements so only the normal state is probed here. We have measured three NMR observables:
 
1) The magnetic hyperfine shift $K$, also called Knight shift, which is calculated from the frequency difference between the measured $^{29}$Si resonance and the calculated resonance of the bare nucleus. $K$ is usually written as the sum of a $T$ independent orbital part and a temperature dependent spin part : $K=K_{\rm orb}+K_{\rm spin}$, where $K_{\rm spin}$ is proportional to the local, uniform $(q=0)$ and static ($\omega=0)$ spin susceptibility $\chi_{\rm spin}$. 

2) The line width, providing a measure of the spatial distribution of hyperfine shifts that may result from chemical, lattice and electronic inhomogeneity on all length scales. In a random powder, the Knight shift anisotropy also contributes to the line width but since the line shape does not show any shoulder or splitting here (Fig.~\ref{NMR}a), the line broadening turns out to be rather dominated by spatial inhomogeneity. The line shape has been fit by a Voigt profile, that is, a convolution of Gaussian and Lorentzian functions of widths $w_G$ and $w_L$, respectively, giving a full width at half maximum FWHM =~$0.5346 \, w_L + \sqrt{0.2166 \, w^2_L+ w^2_G}$. 
 
 3) The spin-lattice relaxation rate, $T_1^{-1}$, providing information on low-energy spin dynamics through the imaginary part of the dynamic spin susceptibility: $(T_1T)^{-1}\propto \sum_q A_\perp^2(q)\chi^{\prime \prime}_{\perp}(q,\omega_n)$, where $\omega_n$ is the nuclear Larmor frequency ($\sim10^8$~Hz here) and $A(q)$ a site-specific hyperfine form factor. $T_1$ values were determined from single-exponential fits of the time dependence of the nuclear magnetization following a saturating pulse. Excellent fits were obtained at all temperatures, without introducing any stretching exponent.

\section{Experimental results and discussion}

\subsection{SMS spectroscopy}\label{MossRes}

\begin{figure}[h]
\includegraphics[width=\linewidth]{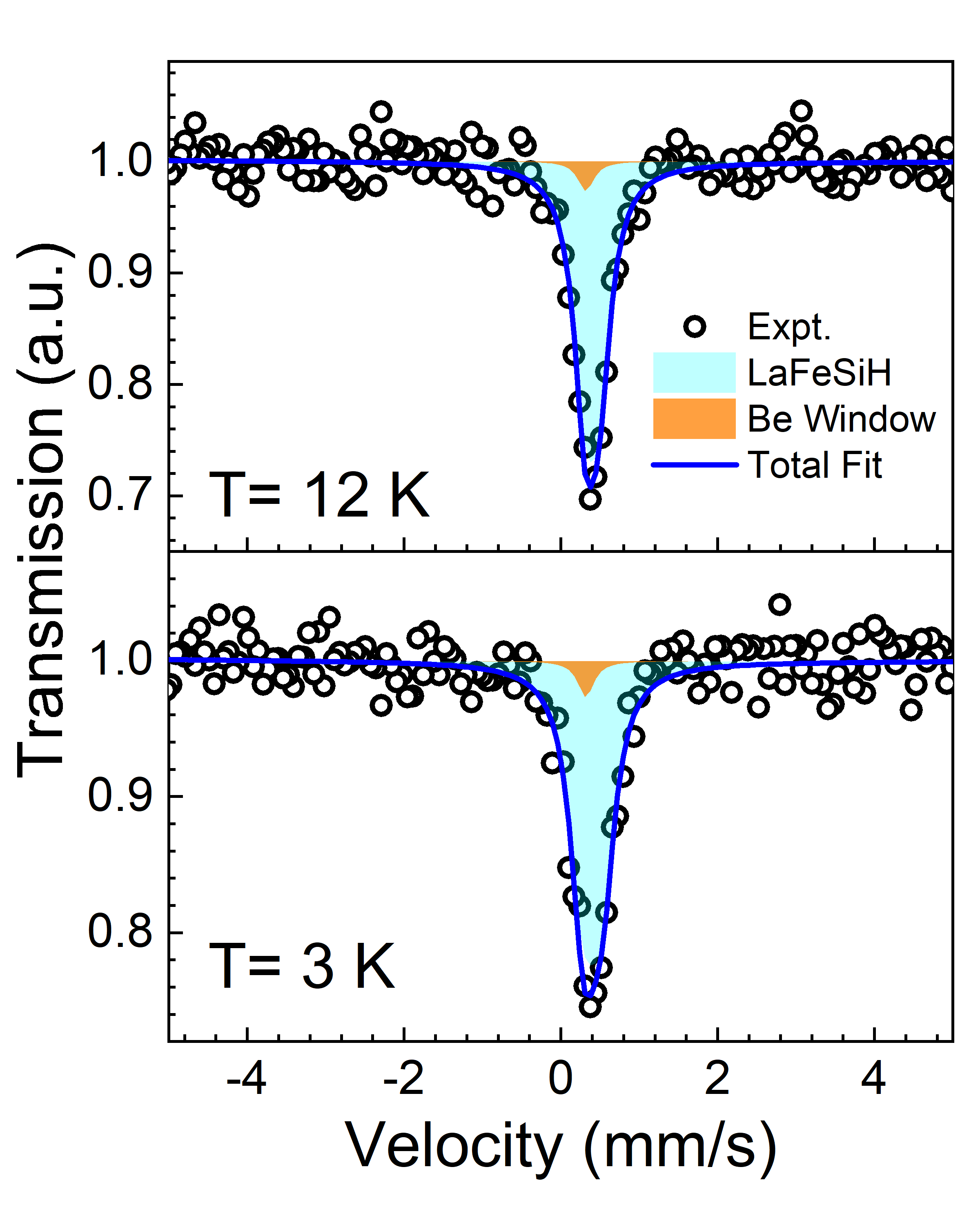}
\caption{\label{Mossb_LT_ambP} (Color online) Synchrotron $^{57}$Fe M\"ossbauer of LaFeSiH natural iron sample at 3 K (superconducting state) and 12 K (normal state) at ambient pressure. Empty circles represent experimental data points. The M\"ossbauer spectra are fitted with single line (cyan fill) plus an additional contribution, also modeled with a single line (orange fill) coming from iron impurities in Beryllium window and collimating lenses of the beamline. The blue solid line through the data points represents the overall fit convoluted with the instrumental resolution. 
}
\end{figure}

In Fig. \ref{Mossb_LT_ambP}, we show SMS spectra of $^{57}$Fe isotope in LaFeSiH taken at low temperature, at 12 K and 3 K, at ambient pressure and without any applied magnetic field. The spectra fitted well by a single line, with possible quadrupolar splitting, at best of the same order of the resolution, of about 0.1 mm/s, also indicating a very weak electric field gradient on the iron site. 

Thus we detect no magnetic hyperfine field on the iron site within the experimental resolution, which for an ordering of Fe magnetic moment, would have been expected to give magnetic splitting. However, a very weak magnetic moment would induce a small splitting, so, a single line shape can not guarantee a zero moment. The instrumental line width and distributions such as isomer shift and quadrupole splitting may also broaden the single line. We estimate an upper bound of the magnetic hyperfine field being $\sim$ 0.6 T.

As a comparison, in LaFeAsO, $^{57}$Fe M{\"o}ssbauer spectroscopy measures a hyperfine magnetic field B$_{hf}$ of 4.86 T, according to Ref. \onlinecite{KlaussPRL.101.077005LaFeAsO}, giving an estimate of the magnetic moment of m(Fe) = 0.25(5) $\mu_B$ at 13 K, while Ref. \onlinecite{KitaoJPSJ2008} and \onlinecite{McGuirePRB.78.094517LaFeAsO} found about 5.3-5.2 T giving a larger estimate of the magnetic moment of about 0.35 $\mu_B$ at a lower temperature of 4.2 K. Despite a difference in the estimation, both values are far above the value we find in LaFeSiH as can be easily seen in the spectra, which clearly shows a magnetic spliting in LaFeAsO at low temperature, contrary to the LaFeSiH case. 

Without a precise knowledge of the electronic and magnetic state of the compound, it is difficult to estimate a upper limit for the magnetic moment in our case, however we can estimate an upper limit of an ordered magnetic moment on iron, in the hypothesis of a similar magnetic structure, below $\sim$ 0.04 $\mu_B$. Indeed, the observed signal is closer to what is measured in tetragonal FeSe \cite{Mizuguchi2010} as well as doped, superconducting LaFeAsO$_{0.89}$F$_{0.11}$ \cite{KitaoJPSJ2008}. 

\begin{figure}[h]
\includegraphics[width=1\linewidth]{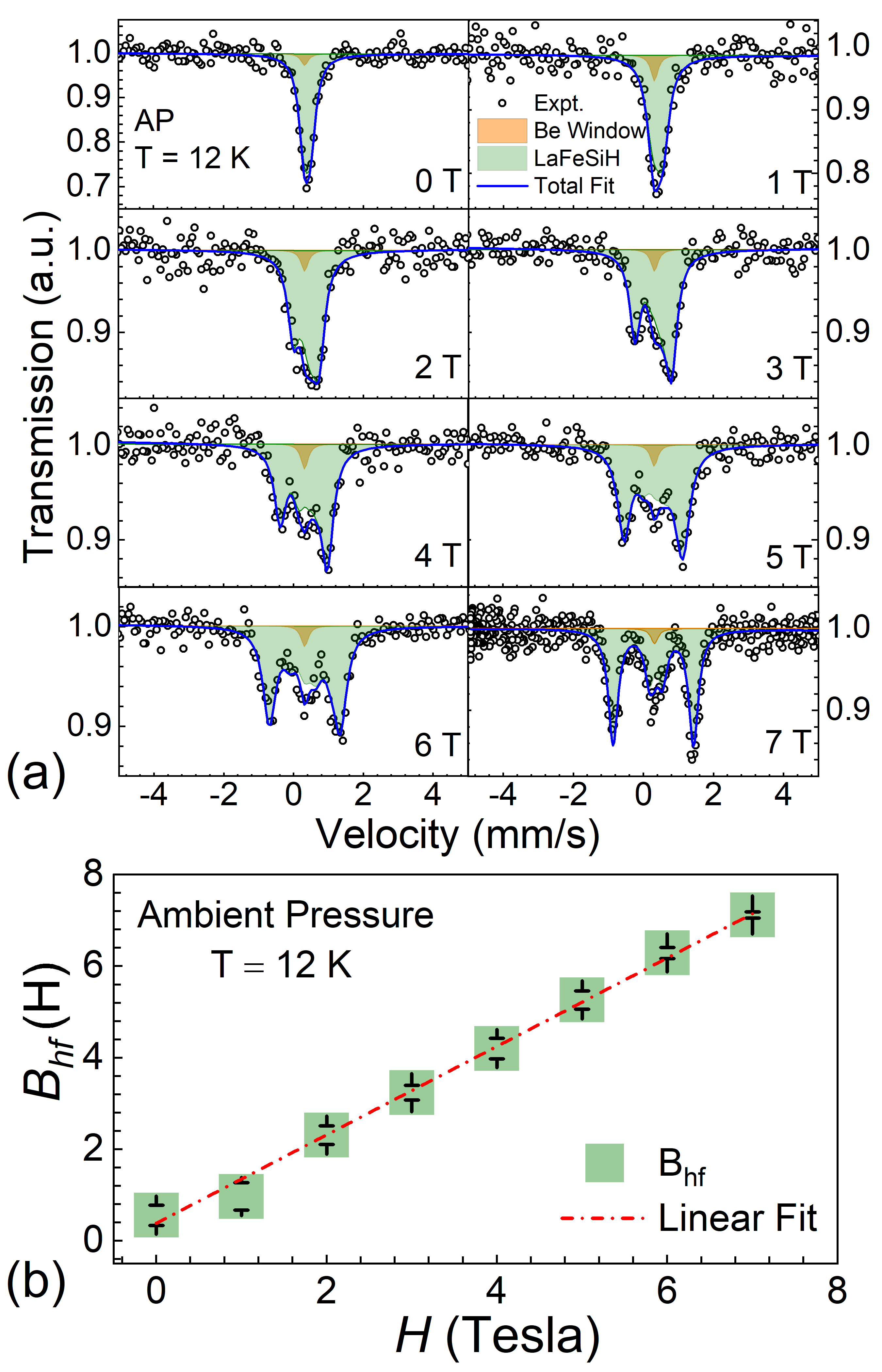}
\caption{\label{LaFeSiH_Field} (Color online) (a) Empty black circles represent SMS  spectra of LaFeSiH natural iron sample at ambient pressure and 12 K at different applied magnetic field from 0 to 7 T whereas the blue solid line through the data points represents the overall fit convoluted with the instrumental resolution. M\"ossbauer spectra were fitted with a magnetic splitting (green) plus a single line, orange filled, coming from iron impurities in Beryllium window and collimating lenses of the beamline. (b) The extracted hyperfine field $B_{hf}(H)$ (green squares), compared to a linear fit (red dot-dashed line).}
\end{figure}

In order to investigate a possible localized, disordered moment, we applied a magnetic field, \textit{H}, up to 7 T. In Fig. \ref{LaFeSiH_Field}, top panel, we show the results ramping up from zero to 7 T for the natural iron sample at 12 K and ambient pressure. Similar results where obtained for the field downstroke on the enriched iron sample. At the maximum, 7 T, field an additional measurement on a second enriched sample gave the same B$_{hf}$ within error bars. In Fig. \ref{LaFeSiH_Field}, bottom panel, we show the fitted B$_{hf}$ as a function of the applied field $H$, fitted by linear regression. The small deviation from the applied field is constant on all the data points for the field upstroke. Our fit gives a slope of 0.98 $\pm$ 0.03, consistent with unity within the error bars, pointing to an effect coming exclusively from the applied field without any contribution from a molecular field acting on the $^{57}$Fe nuclei. A similar observation was made for LaFeAsO$_{0.89}$F$_{0.11}$ \cite{KitaoJPSJ2008}, which also shows an hyperfine field at 4.2 K corresponding within error-bars to the applied external field at 7 T.

\begin{figure}[ht]
\includegraphics[width=1\linewidth]{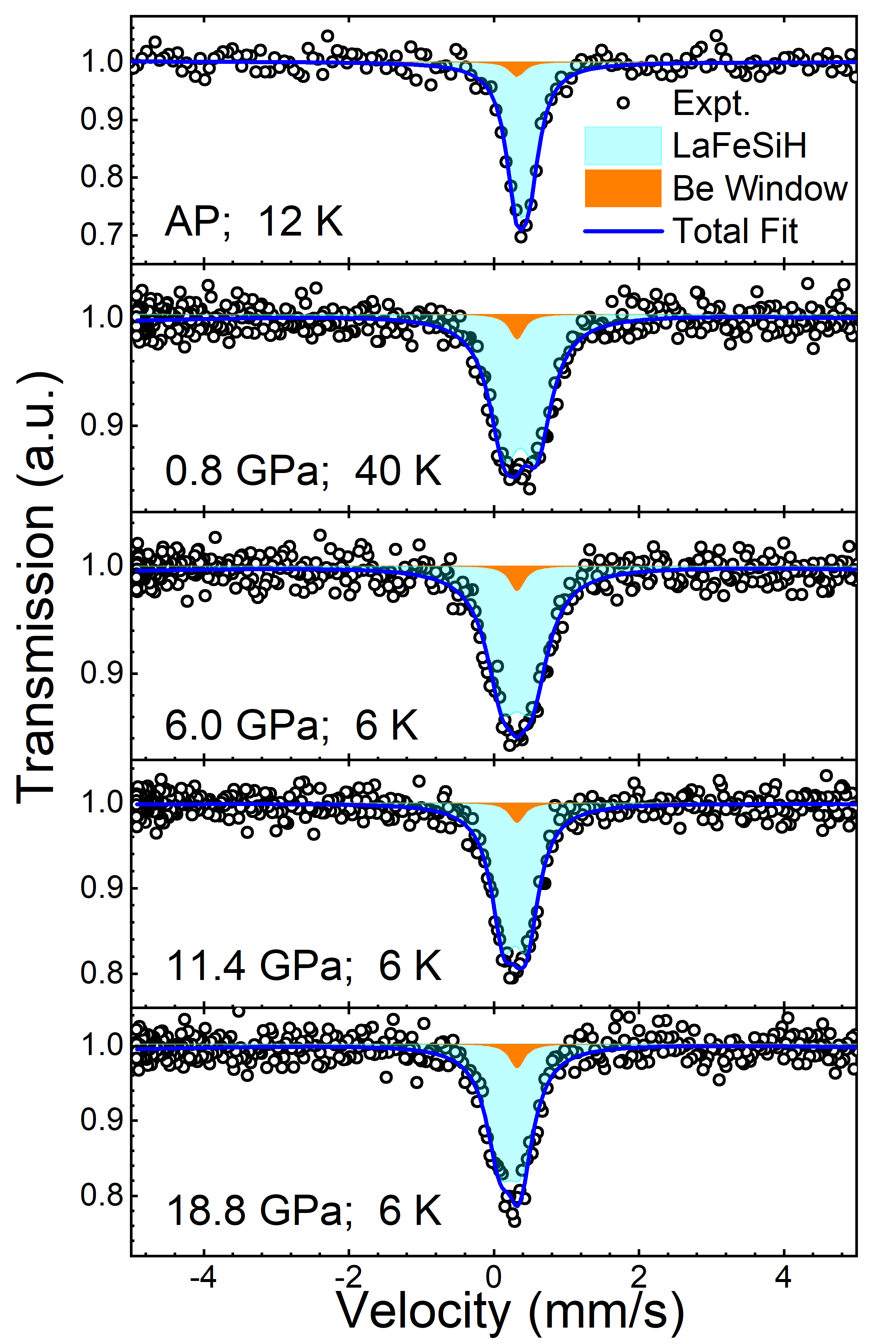}
\caption{\label{HPMoss} Synchrotron $^{57}$Fe M\"ossbauer spectra of LaFeSiH natural iron sample as a function of pressure up to 18.8 GPa, in zero field and  at the lowest temperature for each pressure point. Exact values of pressure and temperature are written in the respective plots. For each pressure, the temperature of the shown data is always above the superconducting one. Note that the data at ambient pressure are taken outside the DAC cell. Same symbols and color code as in Fig. \ref{Mossb_LT_ambP}. 
}
\end{figure}

We note that measurements performed during applied field downstroke on enriched iron sample with 96\% $^{57}$Fe, give results that are perfectly superposed to the upstroke between 5 and 7 T, \textit{i.e.} as long as a magnetic splitting is larger than the intrinsic line-width, which is larger for enriched iron sample because of its broadened signal, compared with natural iron sample. 

We then applied a similar procedure on the same sample, loaded in the high pressure diamond anvil cells as described in Sec. \ref{Mossmeth}. We investigated the M\"ossbauer signal up to 18.8 GPa, and down to 6 K, where we have only performed a zero field measurements, before breaking the diamonds while attempting to reach higher pressure. A summary of the results is shown in Fig. \ref{HPMoss}. 
In LaFeAsO, pressure monotonically decreases both T$_N$ and $B_{hf}$, while initially increasing T$_c$ from zero up to $\sim$ 20 K at about 10 GPa \cite{Kawakami2009_jpsj}. The results shown here in LaFeSiH are similar to the ones at ambient pressure, although a broadening is now observed, that can be fitted with a residual hyperfine field at zero applied field. However, we found that all measurements in the cell give a large broadening of the lineshape, possibly due to vibration of the high-pressure membrane DAC as well as a distribution of the pressure inside the sample. 
In particular, for the highest pressure at zero field,  we estimate a maximum possible hyperfine field of 1.3 $\pm$ 0.3 T, larger than the value found in samples outside the DAC cell, but with a line-shape that can be perfectly interpreted as coming from a single line. 

In summary, the M\"ossbauer data, does not show any splitting of the nuclear resonance which can be linked to a hyperfine magnetic field from a magnetic polarization of the electronic cloud on iron sites, neither at ambient pressure and magnetic field, nor at high pressure, up to $\approx$ 19 GPa. 

Measurements under field give a hyperfine field that corresponds to the applied field within error bars, up to 7 T, both at ambient pressure (Fig. \ref{LaFeSiH_Field}) and up to $\approx$ 12 GPa (see Fig. \ref{HPHFLTMoss} in Appendix \ref{SMSadd}). This rules out a disordered magnetic state with a large, $\approx$ 1 $\mu_B$ moment. 

The present results can not, however, rule out a small moment and/or a very complex (and unusual) order which minimizes the hyperfine field by symmetry.
We note that a very small moment, below m(Fe) $\approx$ 0.05 $\mu_B$, in a disordered state, could be possibly probed using X-ray emission spectroscopy, looking at the K$_\beta$ line satellite, a very sensitive probe to partially filled $3d$ electronic band, as recently shown in $\epsilon$-iron \cite{Lebert20280} and FeSe \cite{Lebert_FeSe}. Another possibility would be to use synchrotron radiation perturbed angular correlation spectroscopy, another approach recently applied to $\epsilon$-iron \cite{PhysRevB.101.035112}.

\subsection{Neutron powder diffraction}
\begin{figure}
    \centering
    \includegraphics[width=\linewidth]{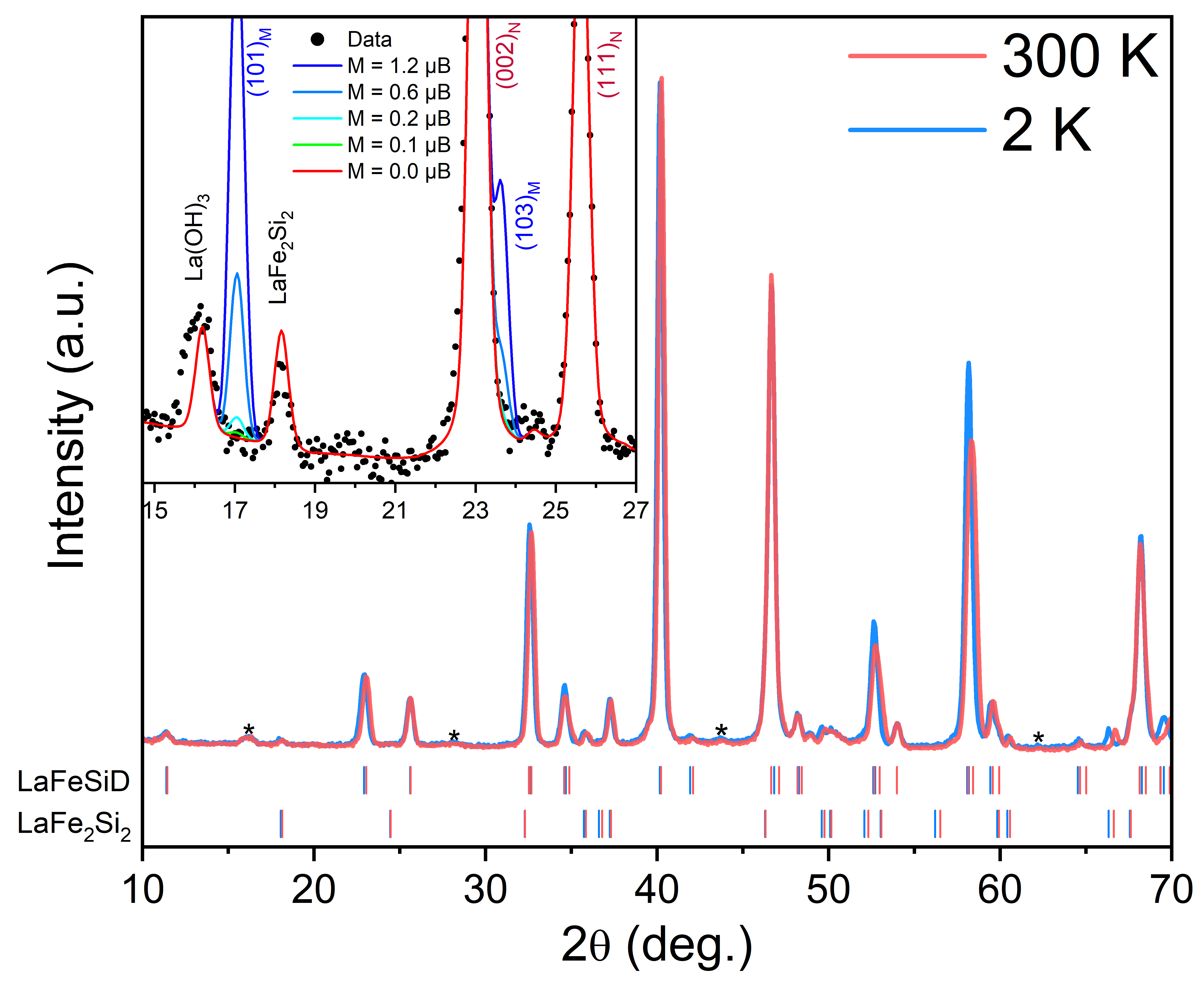}
    \caption{Neutron powder diffraction patterns at 300~K (in red) and 2~K (in blue) recorded at the D2B instrument at ILL. No additional peaks are observed between the two temperatures, consistent with the lack of long range magnetic ordering with \textit{k}$\neq$ 0. Peaks indicated by a star originate form a small La(OH)$_3$ impurity. \textbf{Inset:} Neutron powder diffraction pattern measured at 2~K. Simulations of the single stripe AFM structure is plotted on top of the data, with different magnetic moment values ranging from m(Fe) = 1.2 $\mu_B$ (blue) to m(Fe) = 0.1 $\mu_B$ (green).}
    \label{fig:NPD_data}
\end{figure}

Fig. \ref{fig:NPD_data} shows NPD patterns recorded at 300~K and 2~K. No splitting or additional Bragg peaks are observed at low temperatures, the peaks are only shifted in angle due to the thermal contraction of the unit cell. This is consistent with the absence of a symmetry breaking transition and absence of any long range magnetic order with \textit{k}$\neq$0 at low T. No systematic extra intensity is observed on top of the nuclear diffraction peaks, except changes of relative intensities related to modifications of the atomic positions in the cell. This suggests a lack of long range magnetic ordering related to \textit{k}=0. Thus, our NPD data shows no sign of long range magnetic ordering with a significant magnetic moment m(Fe).

To estimate the detection limit of the NPD experiment we performed several simulations. We used the usual single-stripe-type AFM structure found in LnFeAsO arsenides\cite{delaCruz2008}, with varying moments from m(Fe) = 1.2 $\mu_B$ (i.e. the value obtained from DFT \cite{arribi2020}) to m(Fe) = 0.1 $\mu_B$, taking into account an orthorhombicity of $\delta{} = (b-a)/(b+a)\sim 2~10^{-3}$ i.e. the magnitude measured in DAC at moderate pressures \cite{bernardini-lafesih-prb1}. The calculated pattern of the nuclear model, issued from the Rietveld fit of the observed pattern collected at 2 K, is shown in red in the inset of figure \ref{fig:NPD_data}. By comparing our simulations to the noise level of the data we can estimate that a magnetic order with moments below m(Fe) = $0.15(5) \mu_B$ would not be detected in our experiment. This can thus be considered as an upper bound for m(Fe) from NPD.

\subsection{X-ray powder diffraction}

\begin{figure*}
    \centering
    \includegraphics[width = \textwidth]{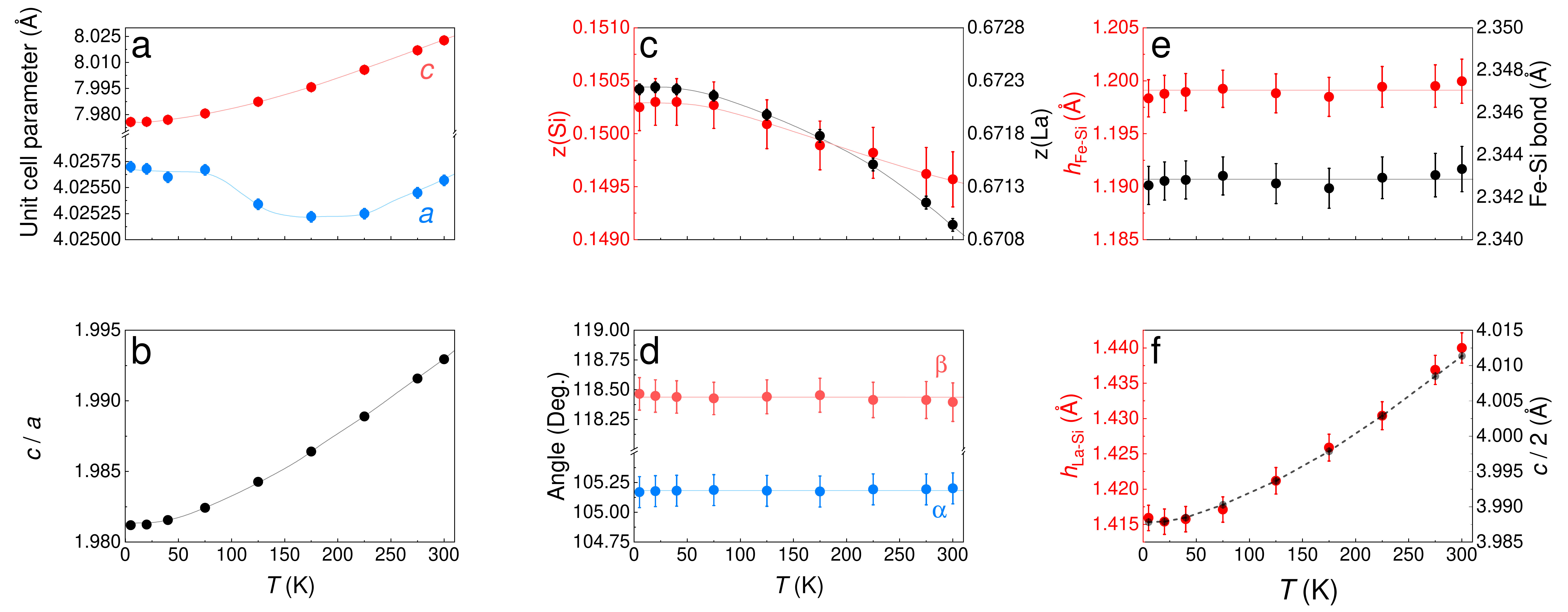}
    \caption{The refined structural parameters as a function of temperature, obtained from synchrotron x-ray diffraction (at the CRISTAL beamline at SOLEIL). \textbf{a:} The unit cell parameter \textit{a} and \textit{c}. \textbf{b:} The unit cell parameter ratio \textit{c}/\textit{a}. \textbf{c:} The fractional z-coordinates of Si and La. \textbf{d:} The FeSi$_4$ tetrahedra angles $\alpha$ and $\beta$. \textbf{e:} The Fe-Si height and bond length. \textbf{f:} In red, the La-Si height. In black, \textit{c}/2. This displays the contraction mechanism upon cooling.}
    \label{fig:LT_XRD_results}
\end{figure*}

In comparison to the NPD experiment, the higher angular resolution of our XRD experiment allows us to confirm that LaFeSiH does not undergo any symmetry lowering structural distortion, as also shown by Raman spectroscopy \cite{layek2023lattice}. This is in contrast to the previously reported tetragonal to orthorhombic distortion detected at 15 K and at moderate pressures \cite{bernardini-lafesih-prb1}. This is however, consistent with the absence of long range magnetic order, since such structural distortion is usually observed before the onset of stripe-type anti-ferromagnetic order in Fe-based superconductors \cite{Qureshi_PRB_2010}.
In figure \ref{fig:LT_XRD_results}.a we show the refined lattice parameters, \textit{a} and \textit{c} respectively, obtained from Rietveld fits of the XRD data using a tetragonal model (\textit{P}4/\textit{nmm}). Firstly, the thermal contraction of the unit cell with temperature is clearly evident for the \textit{c} parameter. The \textit{a} parameter, however, displays an almost static behavior except for a minute broad minimum around 175 K. This minimum, possibly related to the coupling of the lattice with an electronic degree of freedom, was also observed in the refined lattice parameters from NPD (See appendix D). The origin of such behavior can not be readily explained by this data. To uncover the underlying mechanism of this anomaly, further studies are required which are out of the scope of this work.
The refinable atomic positions in the unit cell, z(Si) and z(La), are shown in figure \ref{fig:LT_XRD_results}.c. Both evidence an asymptotic increase with decreasing temperature. This results in the Fe-Si height and Fe-Si bond length remaining nearly unchanged in the entirety of the temperature range, as shown in figure \ref{fig:LT_XRD_results}.e. The relative local environment of the FeSi${_4}$ tetrahedron also remains unmodified, as shown by the temperature dependence of alpha and beta angles in \ref{fig:LT_XRD_results}.d. The contraction of the \textit{c}-axis can be attributed solely to the decrease of the La-Si height, i.e. the inter-block distance. In figure \ref{fig:LT_XRD_results}.f this distance is plotted along with \textit{c}/2 to illustrate the contraction mechanism. 

\subsection{Nuclear magnetic resonance}\label{NMR-part}

\begin{figure*}[t!]
  \centering
    \includegraphics[width=12cm]{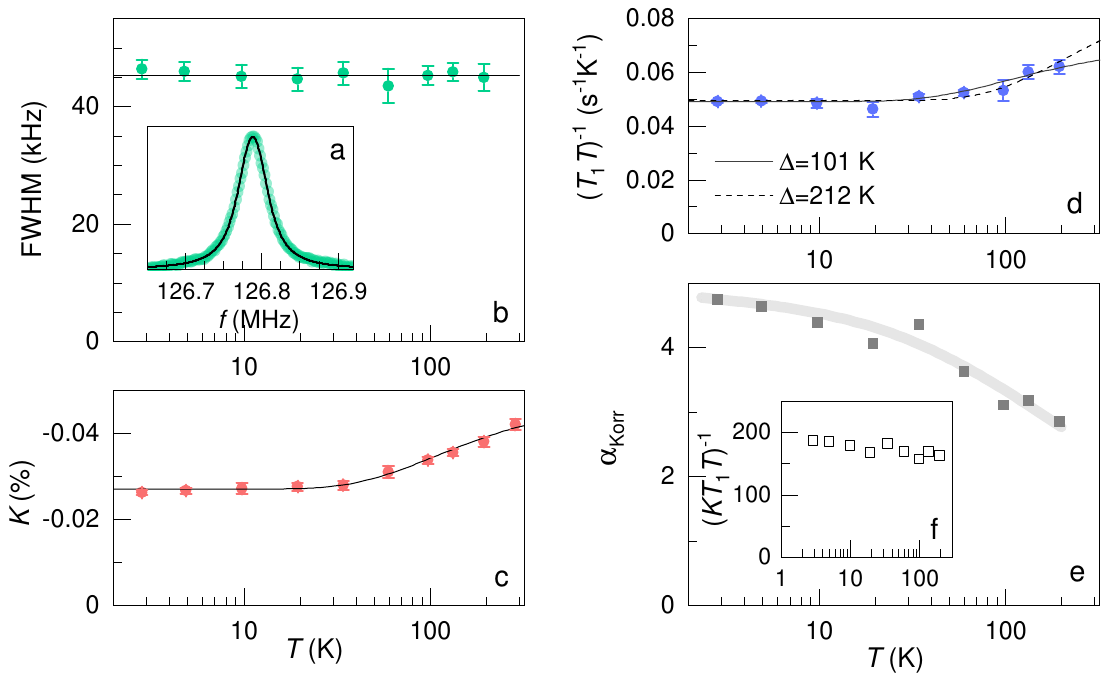}
     \caption{\label{NMR} (a) $^{29}$Si NMR line (green) in a field of $\sim$15~T and $T=2.85$~K. The thin black line is a fit to a Voigt profile. (b) Full width at half maximum (FWHM). The horizontal line is a fit to a constant value 45$\pm$3~kHz. (c) Total magnetic hyperfine shift $K$. The line is a fit to an activated dependence with a gap value $\Delta_K=106\pm24$~K. The negative sign of $K$ with larger absolute values at high temperatures shows that the hyperfine coupling of $^{29}$Si is negative. (d) Spin-lattice relaxation rate $T_1^{-1}$ divided by $T$. The continuous and dashed lines are fits to an activated dependence with gap values $\Delta_{T_1}=101\pm38$~K (reduced $\chi^2=0.61$) and $\Delta_{T_1}=2\Delta_K=212$~K (fixed, reduced $\chi^2=1.33$), respectively.  Although the data does favor $\Delta_{T_1}\simeq\Delta_K$, $\Delta_{T_1}\simeq 2 \Delta_K$, as observed in LaNiAsO$_{1-x}$F$_x$~\cite{Tabuchi}, cannot be fully excluded.  (e) Korringa ratio $\alpha_{\rm Korr}=\hbar\, \gamma_e^2/(4\,\pi \, k_B\, \gamma_n^2) \,(K^2T_1T)^{-1}$. The thick trace is a guide to the eye. (f) The nearly $T$ independent $(KT_1T)^{-1}$.}
\end{figure*}

In general, magnetic order has two main effects on NMR observables. 1) It either shifts, splits or broadens the NMR lines (except in those exceptional cases where the hyperfine field vanishes at the nucleus position for symmetry reasons, which should not be the case for $^{29}$Si here). 2) It produces a sharp change in the spin-lattice relaxation rate $T_1^{-1}$, either an abrupt drop upon entering the ordered state if the transition is first order or a peak at the transition temperature if the transition is second order (and also a peak for a continuous freezing).

None of these features are observed in our $^{29}$Si NMR data in LaFeSiH: all observables measured in our $^{29}$Si NMR data of LaFeSiH show weak and smooth temperature dependence (Fig.~\ref{NMR}), which rules out the presence of magnetic order.

Specifically, the width of the $^{29}$Si NMR line stays constant from room temperature down to 2.8~K (Fig.~\ref{NMR}b). Given that the conjunction of electron correlations and (unavoidable) disorder tends to broaden the NMR lines upon cooling, the constant linewidth here indicates that spin correlations are actually very weak, if any. The absence of stretched-exponential behavior of the magnetization recovery shows that the electronic spin polarization remains spatially homogeneous at all temperatures, which is also consistent with the weakness of spin correlations (even moderate antiferromagnetic correlations in the sister compound LaFeSiO induce stretched-exponential behavior~\cite{Hansen2022}). The constant linewidth here is also consistent with the absence of an orthorhombic transition inferred from XRD (orthorombicity differentiates $K$ values along $a$ and $b$ axes, which splits the NMR lines, see Ref.~\cite{Zhou2020} and references therein). However, the relatively large width in the powder sample obviously makes NMR a much less sensitive probe than XRD in the present case and NMR cannot exclude small and/or short-range distortions due to pinned nematic fluctuations~\cite{Wiecki2021}.

Both the Knight shift (Fig.~\ref{NMR}c) and $(T_1T)^{-1}$ (Fig.~\ref{NMR}d) decrease smoothly upon cooling and saturate to a constant value below $\sim$20~K. A fit to an exponential activation $a+b\times \exp(-\Delta/k_BT)$ yields virtually identical gap values in the Knight shift ($\Delta_K=106\pm24$~K) and in $T_1$ ($\Delta_{T_1}=101\pm38$~K). 
 
At the qualitative level, the exponential dependence of both $K$ and $(T_1T)^{-1}$ over the whole $T$ range is typical of  strongly overdoped Fe-based superconductors~\cite{Yang,Tabuchi,Ma11}. At the quantitative level, our data strikingly resembles results in strongly overdoped Ba(Fe$_{1-x}$Co$_x$)$_2$As$_2$ for which $\Delta_{T_1}\simeq  \Delta_K$ is also observed~\cite{Ning}. The thermally-activated $(T_1T)^{-1}$ has been ascribed to spin fluctuations associated with small momentum transfers (so-called "intra-band" scattering)~\cite{Ning}. Note that this contribution is also present in the underdoped regime but it becomes increasingly masked upon cooling as the growth of $(\pi,0)$ spin fluctuations ("inter-band" scattering) leads to a large upturn in $(T_1T)^{-1}$.~\cite{Ma,Carretta} Such stripe-type spin fluctuations are manifestly absent in LaFeSiH.

Assuming that the hyperfine coupling of $^{29}$Si in LaFeSiH is similar to that of $^{31}$P in BaFe$_2$(As$_{1-x}$P$_x$)$_2$ (0.61~T/$\mu_B$, ref.~\cite{Iye2012}) and considering that our detection limit is set by the linewidth of 45~kHz, an approximate upper bound for an ordered moment is $\sim0.01\mu_B$. This is consistent with the observation that an ordered moment of 0.05~$\mu_B$ in BaFe$_2$(As$_{1-x}$P$_x$)$_2$ ($x\simeq0.3$)~\cite{Hu2015} leads to $T_1$ results for $^{31}$P that are very different~\cite{Hu2015,Iye2012} from those observed here.

The Korringa ratio $\hbar\, \gamma_e^2/(4\,\pi \, k_B\, \gamma_n^2) \,(K^2T_1T)^{-1}=\alpha_{\rm Korr}$ (where $\gamma_e$ and $\gamma_n$ are the electron and nuclear gyromagnetic ratio, respectively) is found to be slightly enhanced with respect to $\alpha_{\rm Korr}=1$ (Fig.~\ref{NMR}e). This points to rather weak electron correlations in LaFeSiH, consistent with the absence of both magnetic order and stripe-type magnetic fluctuations (notice that we implicitly assume here that $K$ is relatively isotropic and that $K_{\rm orb}\ll K_{\rm spin}$ so that $K \simeq K_{\rm spin}$). Nonetheless, $\alpha_{\rm Korr}$ is not constant as a function of $T$, indicating that the Korringa law of Fermi liquids $K_{\rm spin}^2\propto (T_1T)^{-1}$ is not satisfied. Instead, $(K \, T_1T)^{-1}$ is nearly $T$ independent (Fig.~\ref{NMR}f). 

We thus conclude from this NMR data that LaFeSiH probably has a Fermi surface topology close to that of strongly overdoped iron-pnictides, with neither spin order nor even stripe-type fluctuations. This raises the prospect of enhancing the $T_c$ of LaFeSiH by appropriate chemical substitutions. These results contribute to establishing a solid, quantitative phenomenology of NMR data in overdoped Fe-based superconductors that could be confronted to theoretical predictions based on realistic band structures. Another interesting prospect on the theory side is whether {\it ab-initio} calculations are able to account for the negative sign of the hyperfine coupling of $^{29}$Si in LaFeSiH (as shown by the negative values of $K$ in Fig.~\ref{NMR}c). This is opposite to the hyperfine coupling of $^{29}$Si in LaFeSiO, a sister compound with strongly squeezed FeSi layers~\cite{Hansen2022}.

\section{Conclusions}
In summary, from our SMS, NMR and NPD experiments we do not observe any  long range magnetic order or local moments within detection limit in LaFeSiH at ambient pressure, down to 2 K. 
Comparison of our present results using SMS in LaFeSiH with $^{57}$Fe M\"ossbauer literature for Fe-based superconductors\cite{KlaussPRL.101.077005LaFeAsO,McGuirePRB.78.094517LaFeAsO,KitaoJPSJ2008,Mizuguchi2010} suggests a upper bound of 0.04~$\mu_B$ for an ordered magnetic moment, in agreement with NMR which, under similar hypothesis, suggest an upper bound of $\sim$ 0.01~$\mu_B$.
Furthermore, our high-pressure SMS measurements do not indicate any magnetic moment up to 19 GPa and down to 6 K.
From our XRD and NPD experiments we observe no indication of a lattice distortion into an orthorhombic setting down to 2 K, while the resolution of our experiment would allow us to easily detect distortions of the magnitude previously claimed in LaFeSiH \cite{bernardini-lafesih-prb1} or the one observed in the related compound LaFeAsO \cite{delaCruz2008}. 

\begin{acknowledgments}

We thank M.-A. M\'easson, I. Vinograd and R. Zhou for discussions. 

Part of this work was performed at the LNCMI, a member of the European Magnetic Field Laboratory (EMFL).

This work was supported by the ANR-18-CE30-0018-03 Ironman grant.

We acknowledge the European Synchrotron Radiation Facility for provision of synchrotron radiation resources at the Nuclear Resonance beamline (ID18).  

We acknowledge the synchrotron SOLEIL for providing synchrotron radiation resources at the beamline CRISTAL.  

We acknowledge the Institute Laue Langevin (ILL) for providing neutron radiation at the D2B instrument.

\end{acknowledgments}


%

\appendix

\section{SMS specroscopy additional information}\label{SMSadd}

\begin{figure}[h]
\includegraphics[width=0.8\linewidth]{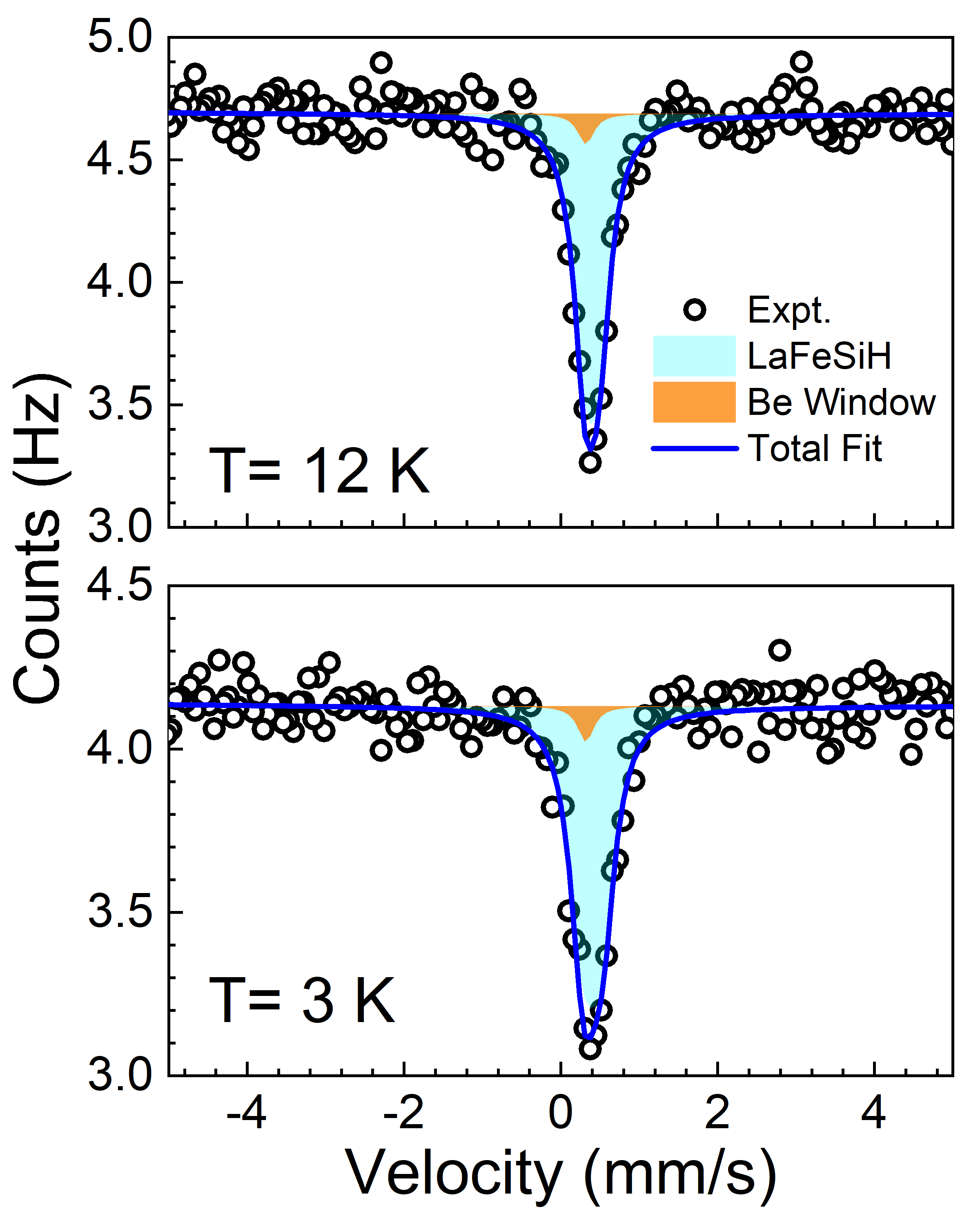}
\caption{\label{Mossb_LT_ambP-raw} (Color online) Synchrotron $^{57}$Fe M\"ossbauer raw data of LaFeSiH natural iron sample at 11.80(5) K (normal state) and 2.96(5) K (superconducting state) at ambient pressure as in Fig.~\ref{Mossb_LT_ambP}. The spectra accumulation times were 1061 sec at 3K and 852 sec at 12 K. Same symbols and color code as in Fig. \ref{Mossb_LT_ambP}. 
}
\end{figure}

\begin{figure}[h]
\includegraphics[width=\linewidth]{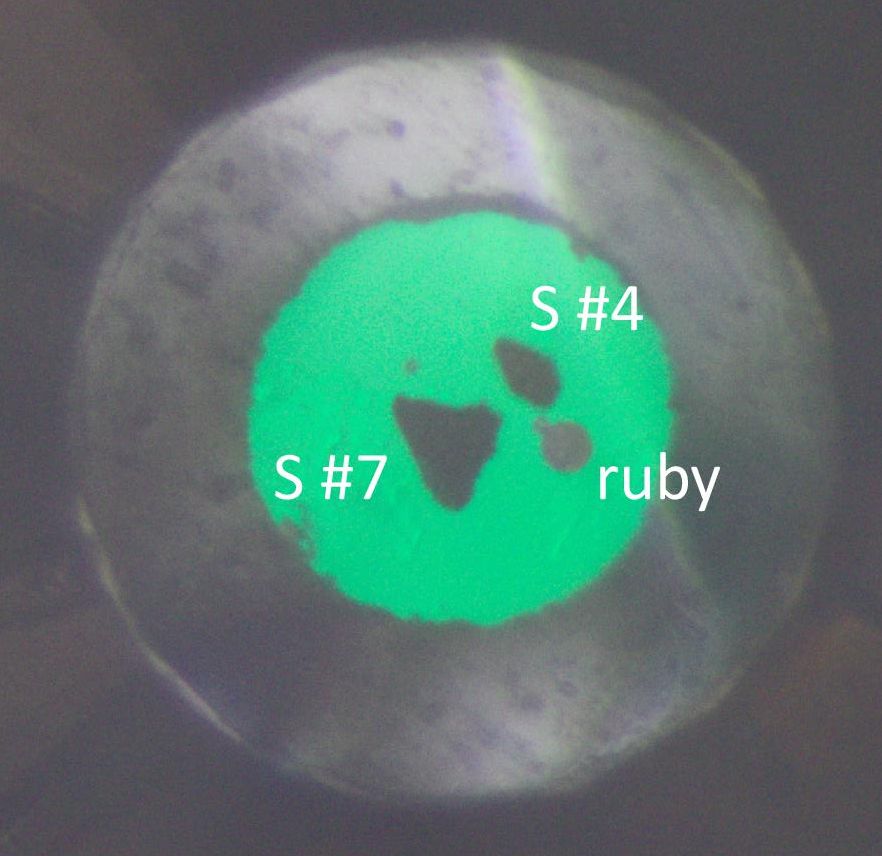}
\caption{\label{LaFeSiHinMDAC} (Color online) Single crystals S \#4 with 96\% $^{57}$Fe substitution and S \#7 with natural iron loaded in the high-pressure cell, together with the ruby pressure gauge. 
}
\end{figure}

We measured 7 crystals at ambient pressure, 2 with natural iron and the remaining 5 highly enriched in $^{57}$Fe. Based on these measurements we selected two of them, one with 96\% $^{57}$Fe substitution and the one with natural iron. 
The central shift and linewidth of the SMS radiation was measured using a single-line K$_2$Mg$^{57}$Fe(CN)$_6$ absorber, typically before and after each measurement point. 
From those measurements the actual center shift and energy resolution of the SMS was derived for the data evaluation of the spectra. Typically, values are 0.395$\pm$0.10 mm/s for the center shift with respect to $\alpha$-Fe and 0.25$\pm$0.05 mm/s for the energy resolution.
The beam from the SMS was focused using Kirkpatrick-Baez Mirror to the spot size of 16 $\mu$m $ \times$ 18 $\mu$m (vertically $ \times$ horizontally, FWHM).  
After a first measurement made on all samples at the lowest temperature (about 3 K), ambient pressure and zero magnetic field, for comparison and reproducibility check, we selected two of them, one with 96\% $^{57}$Fe substitution (hereafter named "enriched iron sample") and the one with natural iron (hereafter named "natural iron sample"), for the remaining experiments in temperature and field. The choice was made on the basis of the smallest width of the line, corresponding to the best quality (\textit{e.g. }lower disorder) of the crystal. 
Samples with 96\% $^{57}$Fe substitution experienced mainly line broadening due to their high effective thickness, while the natural sample gave a resolution limited broadening, although requiring longer accumulation times. 

\begin{figure}[h]
\includegraphics[width=0.8\linewidth]{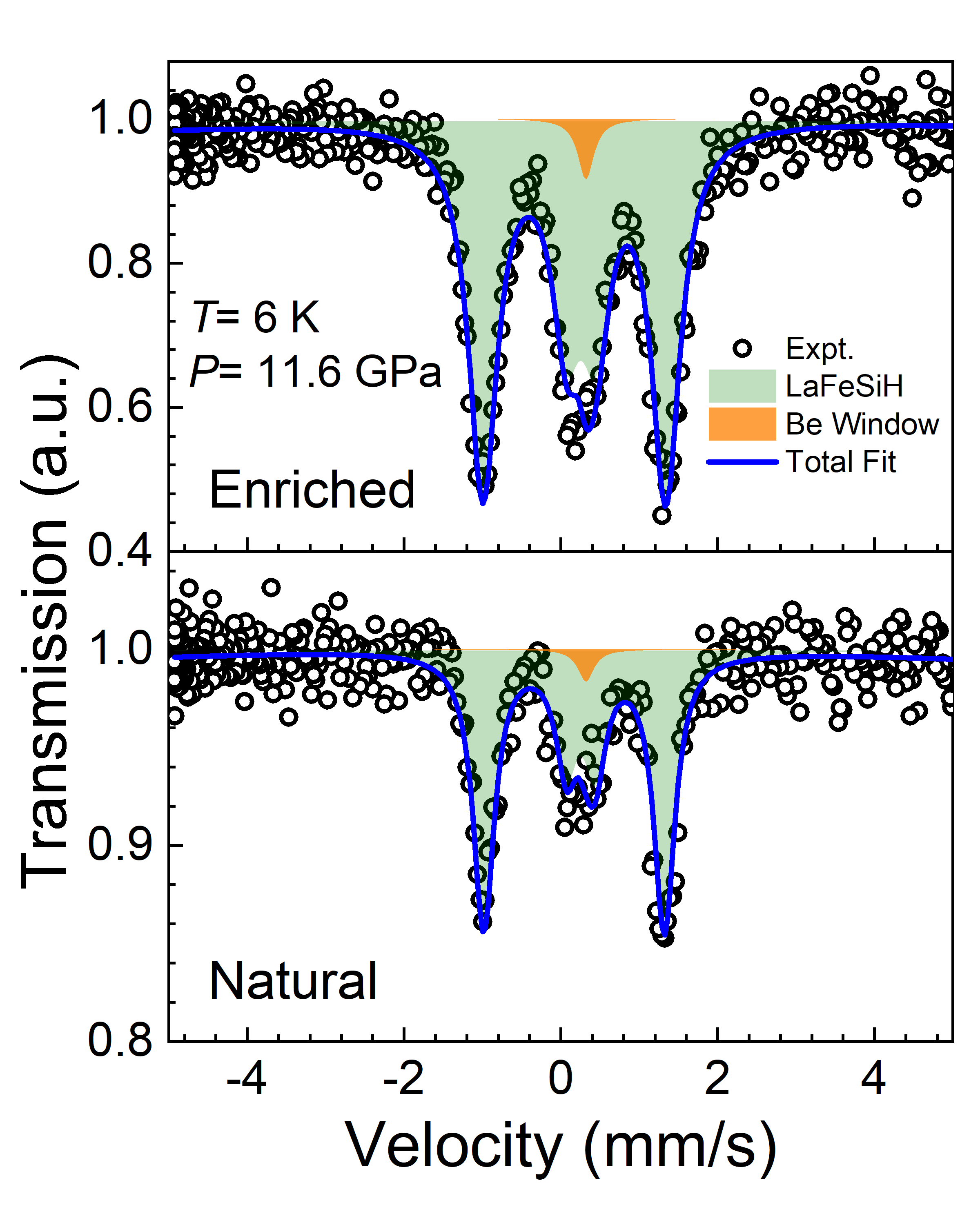}
\caption{\label{HPHFLTMoss} (Color online) SMS spectra of LaFeSiH enriched (top panel) and natural (bottom panel) sample at H=-7 T, T= 6K and P=11.4 GPa. Same symbols and color code as in Fig. \ref{LaFeSiH_Field} panel a). 
}
\end{figure}

In Fig. \ref{Mossb_LT_ambP-raw} we show the raw data of Fig. \ref{Mossb_LT_ambP}, taken with an acquisition time of about 10$^3$ sec. 

In Fig. \ref{LaFeSiHinMDAC} we show a picture of the enriched (S \#4) and natural (S \#7) samples loaded in a DAC with 500 $\mathrm{\mu}$m cullet diamonds and stainless steel gasket was used. The diameter of the gasket hole is 250$\mathrm{\mu}$m, with 70 $\mathrm{\mu}$m thickness chosen to avoid bridging of the single crystal at the highest applied pressure.  

Finally, in Fig. \ref{HPHFLTMoss} we show the measurements obtained at the highest applied field and pressure at low temperature. 
Similar to the measurements at ambient pressure we do not observe any possible magnetic contribution, as the fitted hyperfine field 7.14(6) match the applied one within error bar with a very small residual one. 

\section{Electrical resistivity measurements}\label{resmagmeth}

Resistance of enriched sample as a function of temperature was measured in the four contact geometry down to 4.2 K with an excitation current of 100 $\mu$A using a custom build He$^{4}$ cryostat. Resistance was measured using several other excitation currents in order to check the effect of current on the superconducting transition and normal state resistance behaviour.

\begin{figure}[h]
\includegraphics[width=1.0\linewidth]{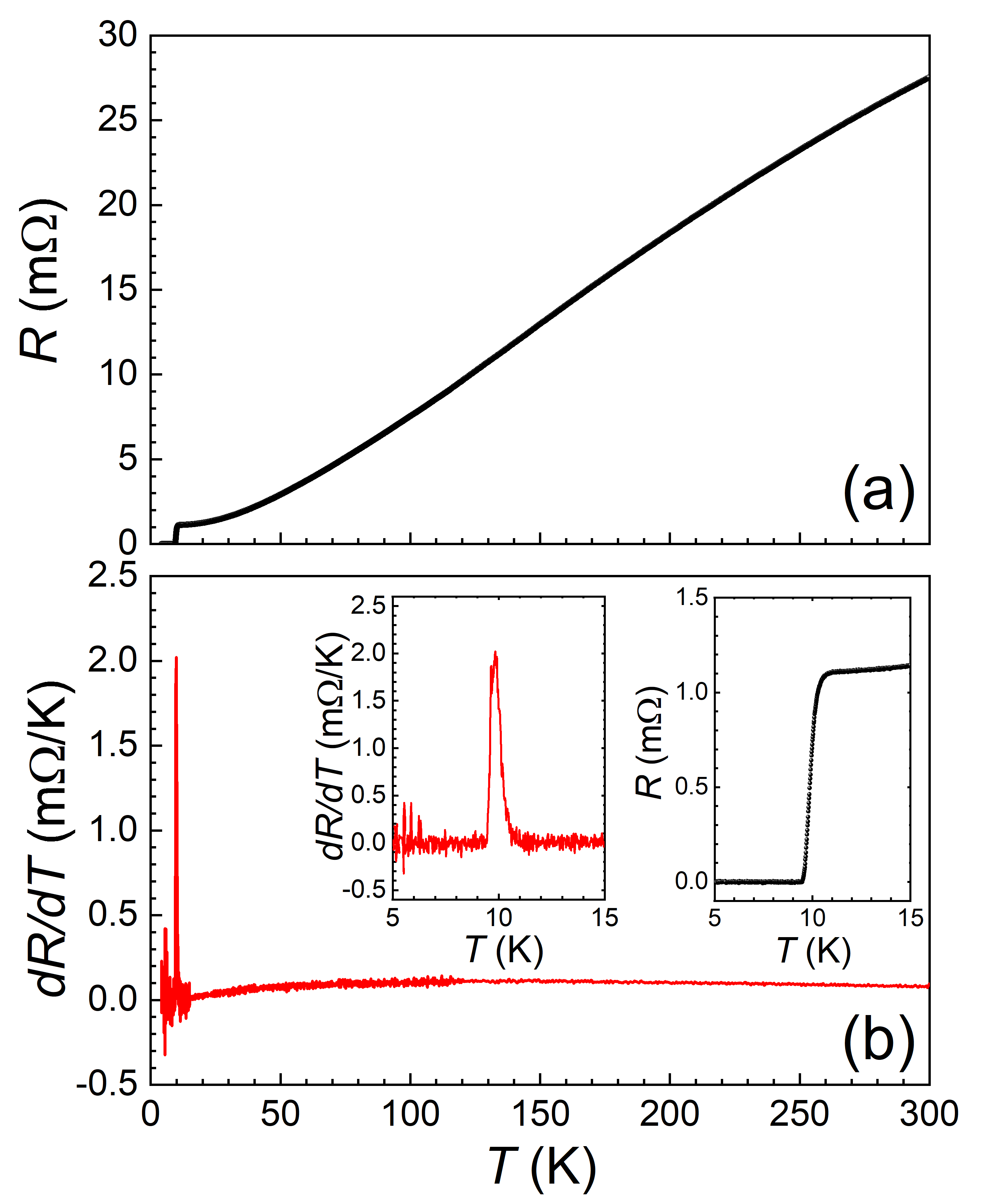}
\caption{\label{R_T_dRdT}(a) Resistance of LaFeSiH single crystal as function of temperature.  (b) Derivative of the resistance as a function of temperature. (Inset) Zoom for temperature range 5-15 K showing superconducting transition around 10 K.
}
\end{figure}

Fig. \ref{R_T_dRdT} (a) shows the measured resistance of LaFeSiH single crystal as function of temperature  which shows a Residual Ratio Resistivity $\frac{R(300K)}{R(11K)}$ of $\sim$ 25. Except the superconducting transition, no anomaly, in particular no change of slope which could suggest the onset of an antiferromagnetic state above the superconducting transition temperature is visible. 

 To further highlight a possible small change of slope, in Fig. \ref{R_T_dRdT}, (b), we plot the derivative of the resistance data as a function of temperature. The very sharp superconducting transition (T$_c^{90\%}$-T$_c^{10\%}$=0.6K) is shown in inset. The T$_c^{Onset}\sim$ 10.5 K is slightly above the average of the whole batch used in M\"ossbauer experiments. However, no other signal is detected, contrary to, \textit{e.g.} LaFeAsO, where a clear maximum in the derivative of the resistance is detected at the anti-ferromagnetic transition, as shown in Ref \onlinecite{KlaussPRL.101.077005LaFeAsO}. No significant difference were found in the normal state resistivity with different applied current.

\section{Magnetization measurements}\label{refinement}

The magnetic properties of the three different samples were probed in a Metronique Ingenierie SQUID magnetometer using the standard extraction method: the deuterated LaFeSiD sample (analysed by neutron powder diffraction and x-ray diffraction) and both LaFeSiH samples made of natural Fe (for NMR and M\"ossbauer experiments) and enriched with Fe$^{57}$ (for M\"ossbauer spectroscopy). Magnetization as a function of applied field (M(H) curves) was measured at fixed temperature (2~K and 150~K). Magnetization versus temperature (M(T) curves) was also acquired at different applied fields (10~Oe, 100~Oe and 1~kOe) after cooling the samples down to 2~K in zero field, measuring during heating (zero field cooled (ZFC) process) and cooling (field cooled (FC) mode) of the sample.

\begin{figure}[h!]
    \centering
    \includegraphics[width = 1.3\linewidth]{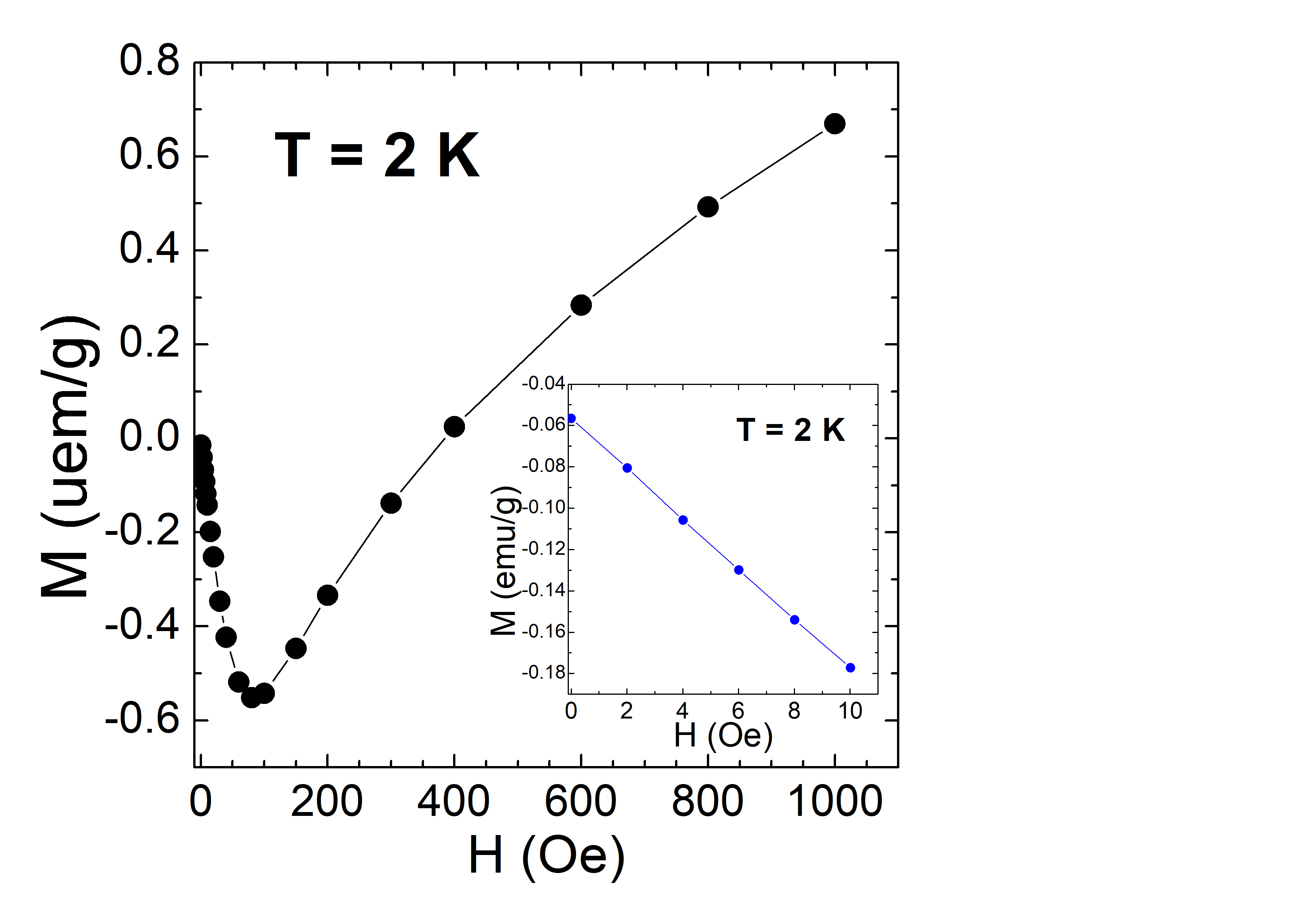}
    \caption{\label{M(H)_LaFeSiD} Magnetic field dependence of LaFeSiD magnetization at T~=~2~K up to 1~kOe. Inset: magnetization at T~=~2~K up to 10 Oe.}
\end{figure}

\begin{figure}[h!]
    \centering
    \includegraphics[width = 1.3\linewidth]{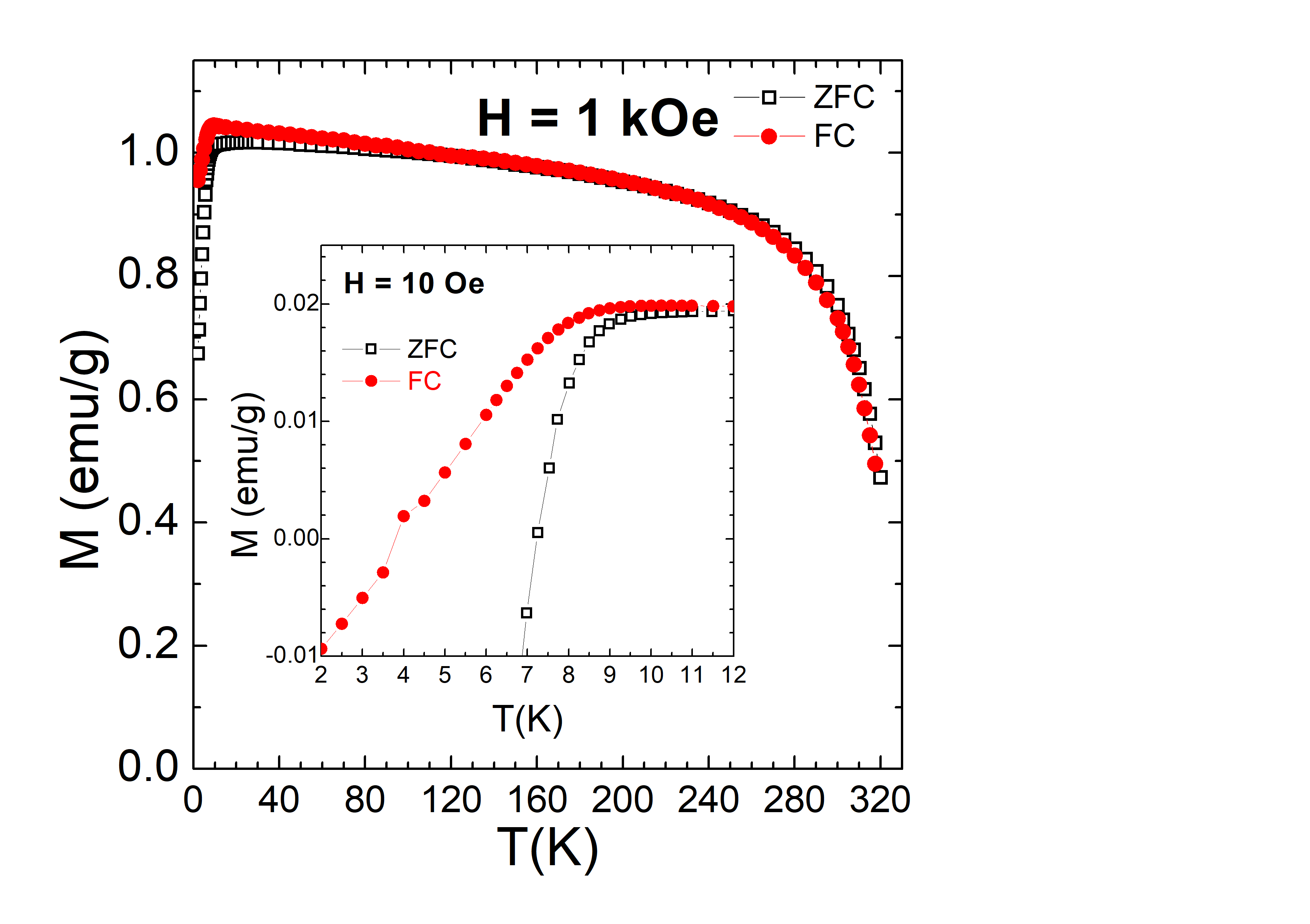}
    \caption{\label{M(T)_LaFeSiD} Temperature dependence of LaFeSiD magnetization at 10~Oe up to 320~K. Inset: enlarged view around T$_c$~=~10~$\pm$~0.2~K.}
\end{figure}

\begin{figure}[h!]
    \centering
    \includegraphics[width = 1.3\linewidth]{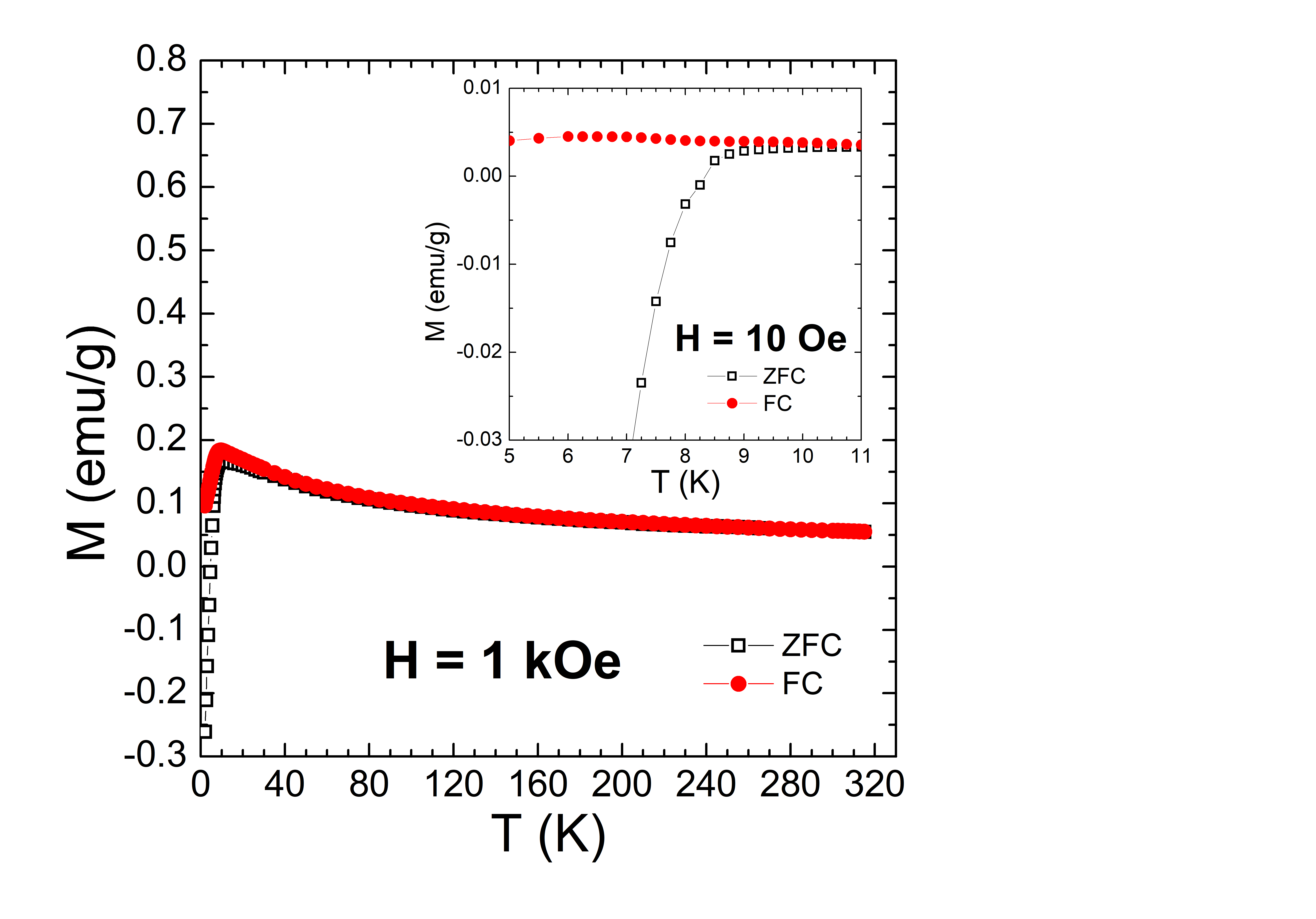}
    \caption{Temperature dependence of isotopic Fe$^{57}$-based LaFeSiH (powder) magnetization at 1~kOe up to 315~K. Inset: enlarged view of M(T) measured at 10~Oe around T$_c$~$\sim$~8.8~K.}
    \label{M(T)_LaFeSiH_57Fe_powder}
\end{figure}

\begin{figure}[h!]
    \centering
    \includegraphics[width = 1.3\linewidth]{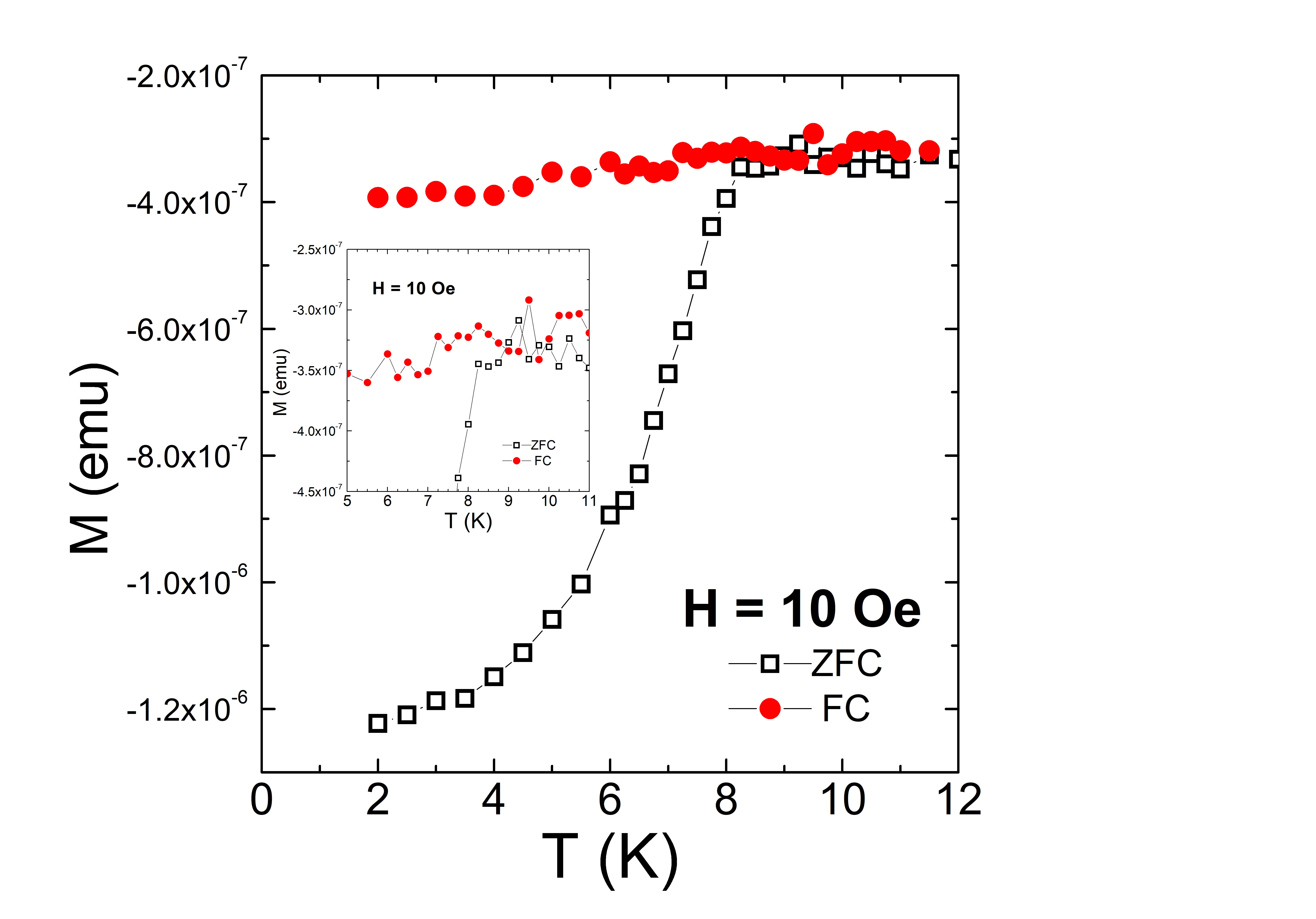}
    \caption{Temperature dependence of isotopic Fe$^{57}$-based LaFeSiH (single crystal) magnetization at 10~Oe in the 2-12~K range. Inset: enlarged view of M(T) measured at 10~Oe around T$_c$.}
    \label{M(T)_LaFeSiH_57Fe_single-crystal}
\end{figure}

The superconducting state is clearly identified in both M(T) and M(H) curves for the different samples, either made from natural Fe or enriched with Fe$^{57}$. The M(H) curve (Fig.~\ref{M(H)_LaFeSiD} for LaFeSiD) shows the expected initial negative slope related to the diamagnetic shielding (at low field, see the linear dependence in the inset of Fig.~\ref{M(H)_LaFeSiD}). From this slope we conclude that superconductivity is bulk in the sample, with a superconducting volume fraction above 80~\% at 2~K for LaFeSiD. Similar or slightly higher values are found for the LaFeSiH samples. The onset of superconductivity (at 10~Oe) is found in the M(T) curve at the critical temperature T$_c$~$\sim$=~10~K for LaFeSiD (inset of Fig.~\ref{M(T)_LaFeSiD}), at T$_c$~$\sim$~8.8~K for the enriched Fe$^{57}$ LaFeSiH batch (inset of Fig.~\ref{M(T)_LaFeSiH_57Fe_powder} for the powder and Fig.~\ref{M(T)_LaFeSiH_57Fe_single-crystal} for the single crystal), and at T$_c$~=~$\sim$~8.5 K for LaFeSiH based on natural Fe (M(T) curve not shown).
All these values are very close to the previously reported T$_c$ values for LaFeSiH powder and single crystal samples\cite{bernardini-lafesih-prb1}.

In the normal state, no significant anomaly is detected, except the ferromagnetic transition associated with the minor deuterated La(Fe$_{1-x}$Si$_x$)$_{13}$D$_y$ phase which presents a Curie temperature near the maximal reached temperature, 320~K\cite{PHEJAR201695,Fujieda} (Fig.~\ref{M(T)_LaFeSiD}). From the saturated magnetization of this impurity, one can estimate that our sample contains less than a weight fraction of $\sim$~2~\% of such phase, which is not detected in our diffraction experiments. The presence of such ferromagnetic background could maybe hide less pronounced transition (like an antiferromagnetic transition) associated with the LaFeSiD main phase. In the Fe$^{57}$-based LaFeSiH, this magnetic background is much smaller (see M(T) curves at 1~kOe, Fig.~\ref{M(T)_LaFeSiD} and Fig.~\ref{M(T)_LaFeSiH_57Fe_powder}) and the signature of potential La(Fe$_{1-x}$Si$_x$)$_{13}$H$_y$ impurity is not detected or its Curie temperature is pushed largely above 315~K. Thanks to this smaller background the Curie-Weiss behavior of the main phase becomes visible and a modelisation by $\chi$~=~$\chi$$_{0}$~+~C/(T-$\theta$$_{W}$) gives $\theta$$_{W}$~$\sim$~-~69~K, highlighting the antiferromagnetic character of the main magnetic interactions in LaFeSiH. We notice also that, except superconductivity, no other anomaly associated with potential structural or electronic/magnetic long range order is directly detectable in the 2~-~315~K range. Other complementary techniques sensitive to magnetism have been used in this work to probe such potential local or long range magnetic order: NPD, M{\"o}ssbauer and NMR spectroscopies.

\section{X-ray and Neutron diffraction - comparison}
The deuterated sample, LaFeSiD was measured using both neutron- and X-ray diffraction.

\begin{table}[ht!]
\caption{The structure of LaFeSiH at T = 300 K and T = 5 K as measured using XRD. \label{structure_param}}
\begin{tabular}{lllllll}
\hline
\multicolumn{7}{l}{P4/nmm}                                                                           \\
\multicolumn{4}{l}{a = 4.02557(1), c = 8.0227(5)} & \multicolumn{3}{l}{a = 4.02570(2), c =7.9756(6)} \\
\multicolumn{4}{l}{T = 300 K}                     & \multicolumn{3}{l}{T = 5 K}                      \\ \hline
Atom       & x        & y        & z              & x            & y            & z                  \\ \hline
La         & 0        & 0.5      & 0.6709(4)      & 0            & 0.5          & 0.6722(4)          \\
Fe         & 0.5      & 0.5      & 0.5            & 0.5          & 0.5          & 0.5                \\
Si         & 0.5      & 0        & 0.1495(7)      & 0.5          & 0            & 0.1502(6)          \\
H          & 0        & 0        & 0              & 0            & 0            & 0                 
\end{tabular}
\end{table}

\begin{figure*}[ht!]
    \centering
    \includegraphics[width = \linewidth]{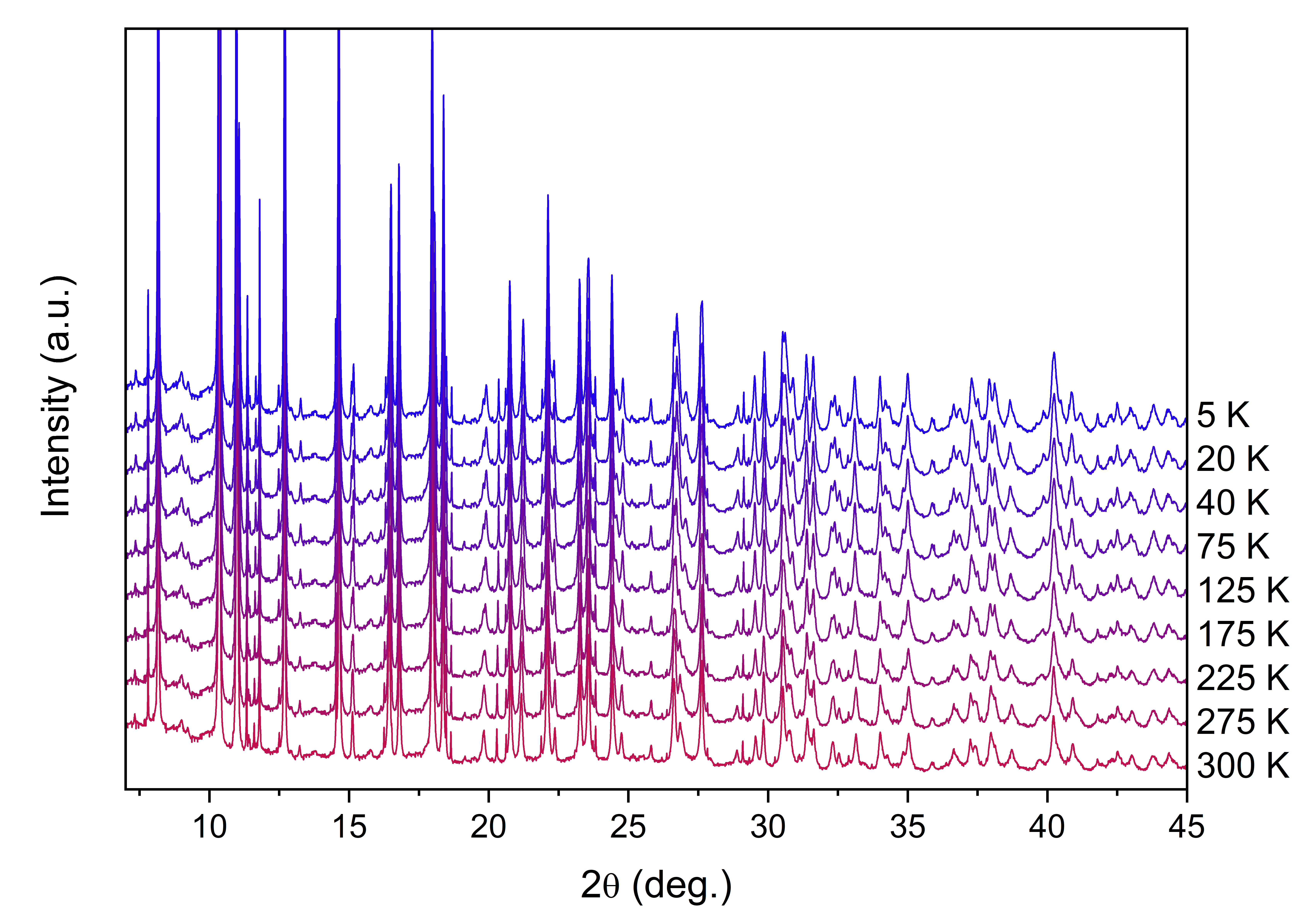}
    \caption{The XRD pattens collected at SOLEIL at different temperatures, using a Mythen 2 detector.}
    \label{XRD_temperature}
\end{figure*}

\begin{figure*}[ht!]
    \centering
    \includegraphics[width = \linewidth]{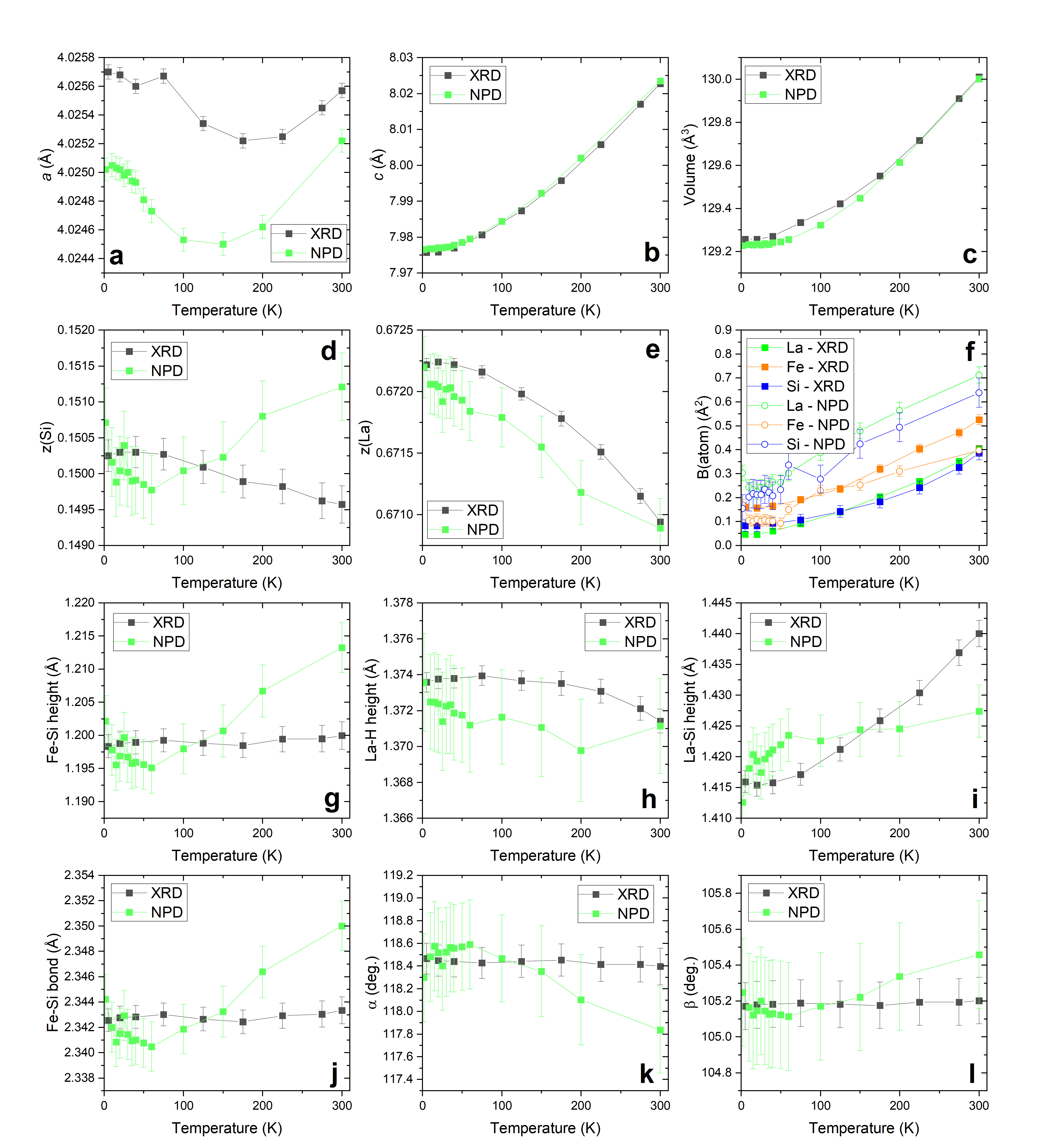}
    \caption{Refined and calculated structural parameters for LaFeSiD as a function of temperature, measured using XRD and NPD. \textbf{a:} The unit cell parameter, \textit{a}. \textbf{b:} The unit cell parameter, \textit{b}. \textbf{c:} The unit cell volume. \textbf{d:} The atomic position parameter z(Si). \textbf{e:} the atomic position parameter z(La). \textbf{f:} The thermal vibration parameters B. For deuterium, the values are in the range of 1.3 for the NPD data and they were not refined for the XRD data. \textbf{g:} The Fe-Si height, along the \textit{c}-axis. \textbf{h:} The La-H(D) height along the \textit{c}-axis. \textbf{i:} The La-Si height along the \textit{c}-axis. \textbf{j:} The Fe-Si bond length. \textbf{k:} $\alpha$ angle of the Fe-Si tetrahedron. \textbf{l:} The $\beta$ angle of the Fe-Si tetrahedron. \label{fig:XRDvsNPD}}
\end{figure*}

In figure \ref{XRD_temperature} we show the XRD patterns collected at various temperatures, showing no splitting og peaks down to 5 K. In figure \ref{fig:XRDvsNPD} we summarize the refined structural parameters and the structural parameters calculated from XRD and NPD. Looking first at figure \ref{fig:XRDvsNPD}.a-c, we show the parameters associated with the unit cell. Here we see a good agreement between the XRD and NPD data, although there is a slight offset. We ascribe this offset to the lack of precision of the neutron wavelength. Then, turning to the atomic position parameters shown in figure \ref{fig:XRDvsNPD}.d-f, we see that the z(Si) parameter is not excellently matched between the two experiments. This can be explained by the rather low scattering cross-section of Si with neutrons, and we thus find it suitable to trust the parameters obtained from XRD. For the z(La) parameter we find a decent match between the two experiments. The thermal parameters, B, all show low positive values, decreasing as temperature decreases, indicative of a reliable fit. In figure \ref{fig:XRDvsNPD}.g-l we show various geometric parameters, calculated from the refined parameters. We see that the values associated with Si, show some deviation between the XRD and NPD data, as we could expect given the above argument, however, the general trends are well matched. Table \ref{structure_param} summarizes the structure at 300 K.

\newpage

\begin{thebibliography}{43}%
\makeatletter
\providecommand \@ifxundefined [1]{%
 \@ifx{#1\undefined}
}%
\providecommand \@ifnum [1]{%
 \ifnum #1\expandafter \@firstoftwo
 \else \expandafter \@secondoftwo
 \fi
}%
\providecommand \@ifx [1]{%
 \ifx #1\expandafter \@firstoftwo
 \else \expandafter \@secondoftwo
 \fi
}%
\providecommand \natexlab [1]{#1}%
\providecommand \enquote  [1]{``#1''}%
\providecommand \bibnamefont  [1]{#1}%
\providecommand \bibfnamefont [1]{#1}%
\providecommand \citenamefont [1]{#1}%
\providecommand \href@noop [0]{\@secondoftwo}%
\providecommand \href [0]{\begingroup \@sanitize@url \@href}%
\providecommand \@href[1]{\@@startlink{#1}\@@href}%
\providecommand \@@href[1]{\endgroup#1\@@endlink}%
\providecommand \@sanitize@url [0]{\catcode `\\12\catcode `\$12\catcode
  `\&12\catcode `\#12\catcode `\^12\catcode `\_12\catcode `\%12\relax}%
\providecommand \@@startlink[1]{}%
\providecommand \@@endlink[0]{}%
\providecommand \url  [0]{\begingroup\@sanitize@url \@url }%
\providecommand \@url [1]{\endgroup\@href {#1}{\urlprefix }}%
\providecommand \urlprefix  [0]{URL }%
\providecommand \Eprint [0]{\href }%
\providecommand \doibase [0]{http://dx.doi.org/}%
\providecommand \selectlanguage [0]{\@gobble}%
\providecommand \bibinfo  [0]{\@secondoftwo}%
\providecommand \bibfield  [0]{\@secondoftwo}%
\providecommand \translation [1]{[#1]}%
\providecommand \BibitemOpen [0]{}%
\providecommand \bibitemStop [0]{}%
\providecommand \bibitemNoStop [0]{.\EOS\space}%
\providecommand \EOS [0]{\spacefactor3000\relax}%
\providecommand \BibitemShut  [1]{\csname bibitem#1\endcsname}%
\let\auto@bib@innerbib\@empty
\bibitem [{\citenamefont {Hosono}\ and\ \citenamefont
  {Kuroki}(2015)}]{hosono-15}%
  \BibitemOpen
  \bibfield  {author} {\bibinfo {author} {\bibfnamefont {H.}~\bibnamefont
  {Hosono}}\ and\ \bibinfo {author} {\bibfnamefont {K.}~\bibnamefont
  {Kuroki}},\ }\href {\doibase https://doi.org/10.1016/j.physc.2015.02.020}
  {\bibfield  {journal} {\bibinfo  {journal} {Physica C: Superconductivity and
  its Applications}\ }\textbf {\bibinfo {volume} {514}},\ \bibinfo {pages} {399
  } (\bibinfo {year} {2015})}\BibitemShut {NoStop}%
\bibitem [{\citenamefont {Martinelli}\ \emph {et~al.}(2016)\citenamefont
  {Martinelli}, \citenamefont {Bernardini},\ and\ \citenamefont
  {Massidda}}]{MARTINELLI20165}%
  \BibitemOpen
  \bibfield  {author} {\bibinfo {author} {\bibfnamefont {A.}~\bibnamefont
  {Martinelli}}, \bibinfo {author} {\bibfnamefont {F.}~\bibnamefont
  {Bernardini}}, \ and\ \bibinfo {author} {\bibfnamefont {S.}~\bibnamefont
  {Massidda}},\ }\href {\doibase https://doi.org/10.1016/j.crhy.2015.06.001}
  {\bibfield  {journal} {\bibinfo  {journal} {Comptes Rendus Physique}\
  }\textbf {\bibinfo {volume} {17}},\ \bibinfo {pages} {5 } (\bibinfo {year}
  {2016})}\BibitemShut {NoStop}%
\bibitem [{\citenamefont {Iimura}\ and\ \citenamefont
  {Hosono}(2020)}]{hosono-20}%
  \BibitemOpen
  \bibfield  {author} {\bibinfo {author} {\bibfnamefont {S.}~\bibnamefont
  {Iimura}}\ and\ \bibinfo {author} {\bibfnamefont {H.}~\bibnamefont
  {Hosono}},\ }\href {\doibase 10.7566/JPSJ.89.051006} {\bibfield  {journal}
  {\bibinfo  {journal} {Journal of the Physical Society of Japan}\ }\textbf
  {\bibinfo {volume} {89}},\ \bibinfo {pages} {051006} (\bibinfo {year}
  {2020})}\BibitemShut {NoStop}%
\bibitem [{\citenamefont {Bernardini}\ \emph {et~al.}(2018)\citenamefont
  {Bernardini}, \citenamefont {Garbarino}, \citenamefont {Sulpice},
  \citenamefont {N\`u{\~n}ez-Regueiro}, \citenamefont {Gaudin}, \citenamefont
  {Chevalier}, \citenamefont {M\'easson}, \citenamefont {Cano},\ and\
  \citenamefont {Tenc\'e}}]{bernardini-lafesih-prb1}%
  \BibitemOpen
  \bibfield  {author} {\bibinfo {author} {\bibfnamefont {F.}~\bibnamefont
  {Bernardini}}, \bibinfo {author} {\bibfnamefont {G.}~\bibnamefont
  {Garbarino}}, \bibinfo {author} {\bibfnamefont {A.}~\bibnamefont {Sulpice}},
  \bibinfo {author} {\bibfnamefont {M.}~\bibnamefont {N\`u{\~n}ez-Regueiro}},
  \bibinfo {author} {\bibfnamefont {E.}~\bibnamefont {Gaudin}}, \bibinfo
  {author} {\bibfnamefont {B.}~\bibnamefont {Chevalier}}, \bibinfo {author}
  {\bibfnamefont {M.-A.}\ \bibnamefont {M\'easson}}, \bibinfo {author}
  {\bibfnamefont {A.}~\bibnamefont {Cano}}, \ and\ \bibinfo {author}
  {\bibfnamefont {S.}~\bibnamefont {Tenc\'e}},\ }\href {\doibase
  10.1103/PhysRevB.97.100504} {\bibfield  {journal} {\bibinfo  {journal} {Phys.
  Rev. B}\ }\textbf {\bibinfo {volume} {97}},\ \bibinfo {pages} {100504(R)}
  (\bibinfo {year} {2018})}\BibitemShut {NoStop}%
\bibitem [{\citenamefont {Vaney}\ \emph {et~al.}(2022)\citenamefont {Vaney},
  \citenamefont {Vignolle}, \citenamefont {Demourgues}, \citenamefont {Gaudin},
  \citenamefont {Durand}, \citenamefont {Labrugère}, \citenamefont
  {Bernardini}, \citenamefont {Cano},\ and\ \citenamefont {Tencé}}]{LaFeSiF}%
  \BibitemOpen
  \bibfield  {author} {\bibinfo {author} {\bibfnamefont {J.-B.}\ \bibnamefont
  {Vaney}}, \bibinfo {author} {\bibfnamefont {B.}~\bibnamefont {Vignolle}},
  \bibinfo {author} {\bibfnamefont {A.}~\bibnamefont {Demourgues}}, \bibinfo
  {author} {\bibfnamefont {E.}~\bibnamefont {Gaudin}}, \bibinfo {author}
  {\bibfnamefont {E.}~\bibnamefont {Durand}}, \bibinfo {author} {\bibfnamefont
  {C.}~\bibnamefont {Labrugère}}, \bibinfo {author} {\bibfnamefont
  {F.}~\bibnamefont {Bernardini}}, \bibinfo {author} {\bibfnamefont
  {A.}~\bibnamefont {Cano}}, \ and\ \bibinfo {author} {\bibfnamefont
  {S.}~\bibnamefont {Tencé}},\ }\href {\doibase 10.1038/s41467-022-29043-8}
  {\bibfield  {journal} {\bibinfo  {journal} {Nat Commun}\ }\textbf {\bibinfo
  {volume} {13}},\ \bibinfo {pages} {1462} (\bibinfo {year}
  {2022})}\BibitemShut {NoStop}%
\bibitem [{\citenamefont {Hansen}\ \emph {et~al.}(2022)\citenamefont {Hansen},
  \citenamefont {Vaney}, \citenamefont {Lepoittevin}, \citenamefont
  {Bernardini}, \citenamefont {Gaudin}, \citenamefont {Nassif}, \citenamefont
  {M{\'e}asson}, \citenamefont {Sulpice}, \citenamefont {Mayaffre},
  \citenamefont {Julien}, \citenamefont {Tenc{\'e}}, \citenamefont {Cano},\
  and\ \citenamefont {Toulemonde}}]{Hansen2022}%
  \BibitemOpen
  \bibfield  {author} {\bibinfo {author} {\bibfnamefont {M.~F.}\ \bibnamefont
  {Hansen}}, \bibinfo {author} {\bibfnamefont {J.-B.}\ \bibnamefont {Vaney}},
  \bibinfo {author} {\bibfnamefont {C.}~\bibnamefont {Lepoittevin}}, \bibinfo
  {author} {\bibfnamefont {F.}~\bibnamefont {Bernardini}}, \bibinfo {author}
  {\bibfnamefont {E.}~\bibnamefont {Gaudin}}, \bibinfo {author} {\bibfnamefont
  {V.}~\bibnamefont {Nassif}}, \bibinfo {author} {\bibfnamefont {M.-A.}\
  \bibnamefont {M{\'e}asson}}, \bibinfo {author} {\bibfnamefont
  {A.}~\bibnamefont {Sulpice}}, \bibinfo {author} {\bibfnamefont
  {H.}~\bibnamefont {Mayaffre}}, \bibinfo {author} {\bibfnamefont {M.-H.}\
  \bibnamefont {Julien}}, \bibinfo {author} {\bibfnamefont {S.}~\bibnamefont
  {Tenc{\'e}}}, \bibinfo {author} {\bibfnamefont {A.}~\bibnamefont {Cano}}, \
  and\ \bibinfo {author} {\bibfnamefont {P.}~\bibnamefont {Toulemonde}},\
  }\href {\doibase 10.1038/s41535-022-00493-z} {\bibfield  {journal} {\bibinfo
  {journal} {npj Quantum Materials}\ }\textbf {\bibinfo {volume} {7}},\
  \bibinfo {pages} {86} (\bibinfo {year} {2022})}\BibitemShut {NoStop}%
\bibitem [{\citenamefont {Hung}\ and\ \citenamefont
  {Yildirim}(2018)}]{yildirim18-prb}%
  \BibitemOpen
  \bibfield  {author} {\bibinfo {author} {\bibfnamefont {L.}~\bibnamefont
  {Hung}}\ and\ \bibinfo {author} {\bibfnamefont {T.}~\bibnamefont
  {Yildirim}},\ }\href {\doibase 10.1103/PhysRevB.97.224501} {\bibfield
  {journal} {\bibinfo  {journal} {Phys. Rev. B}\ }\textbf {\bibinfo {volume}
  {97}},\ \bibinfo {pages} {224501} (\bibinfo {year} {2018})}\BibitemShut
  {NoStop}%
\bibitem [{\citenamefont {Bhattacharyya}\ \emph {et~al.}(2020)\citenamefont
  {Bhattacharyya}, \citenamefont {Rodi\`ere}, \citenamefont {Vaney},
  \citenamefont {Biswas}, \citenamefont {Hillier}, \citenamefont {Bosin},
  \citenamefont {Bernardini}, \citenamefont {Tenc\'e}, \citenamefont {Adroja},\
  and\ \citenamefont {Cano}}]{bhattacharyya2019}%
  \BibitemOpen
  \bibfield  {author} {\bibinfo {author} {\bibfnamefont {A.}~\bibnamefont
  {Bhattacharyya}}, \bibinfo {author} {\bibfnamefont {P.}~\bibnamefont
  {Rodi\`ere}}, \bibinfo {author} {\bibfnamefont {J.-B.}\ \bibnamefont
  {Vaney}}, \bibinfo {author} {\bibfnamefont {P.~K.}\ \bibnamefont {Biswas}},
  \bibinfo {author} {\bibfnamefont {A.~D.}\ \bibnamefont {Hillier}}, \bibinfo
  {author} {\bibfnamefont {A.}~\bibnamefont {Bosin}}, \bibinfo {author}
  {\bibfnamefont {F.}~\bibnamefont {Bernardini}}, \bibinfo {author}
  {\bibfnamefont {S.}~\bibnamefont {Tenc\'e}}, \bibinfo {author} {\bibfnamefont
  {D.~T.}\ \bibnamefont {Adroja}}, \ and\ \bibinfo {author} {\bibfnamefont
  {A.}~\bibnamefont {Cano}},\ }\href {\doibase 10.1103/PhysRevB.101.224502}
  {\bibfield  {journal} {\bibinfo  {journal} {Phys. Rev. B}\ }\textbf {\bibinfo
  {volume} {101}},\ \bibinfo {pages} {224502} (\bibinfo {year}
  {2020})}\BibitemShut {NoStop}%
\bibitem [{\citenamefont {{P. Villar Arribi}}\ \emph
  {et~al.}(2020)\citenamefont {{P. Villar Arribi}}, \citenamefont {Bernardini},
  \citenamefont {de' Medici}, \citenamefont {Toulemonde}, \citenamefont
  {Tenc{\'{e}}},\ and\ \citenamefont {Cano}}]{arribi2020}%
  \BibitemOpen
  \bibfield  {author} {\bibinfo {author} {\bibnamefont {{P. Villar Arribi}}},
  \bibinfo {author} {\bibfnamefont {F.}~\bibnamefont {Bernardini}}, \bibinfo
  {author} {\bibfnamefont {L.}~\bibnamefont {de' Medici}}, \bibinfo {author}
  {\bibfnamefont {P.}~\bibnamefont {Toulemonde}}, \bibinfo {author}
  {\bibfnamefont {S.}~\bibnamefont {Tenc{\'{e}}}}, \ and\ \bibinfo {author}
  {\bibfnamefont {A.}~\bibnamefont {Cano}},\ }\href {\doibase
  10.1209/0295-5075/128/47004} {\bibfield  {journal} {\bibinfo  {journal}
  {{EPL} (Europhysics Letters)}\ }\textbf {\bibinfo {volume} {128}},\ \bibinfo
  {pages} {47004} (\bibinfo {year} {2020})}\BibitemShut {NoStop}%
\bibitem [{\citenamefont {Klauss}\ \emph {et~al.}(2008)\citenamefont {Klauss},
  \citenamefont {Luetkens}, \citenamefont {Klingeler}, \citenamefont {Hess},
  \citenamefont {Litterst}, \citenamefont {Kraken}, \citenamefont {Korshunov},
  \citenamefont {Eremin}, \citenamefont {Drechsler}, \citenamefont {Khasanov},
  \citenamefont {Amato}, \citenamefont {Hamann-Borrero}, \citenamefont {Leps},
  \citenamefont {Kondrat}, \citenamefont {Behr}, \citenamefont {Werner},\ and\
  \citenamefont {B\"uchner}}]{KlaussPRL.101.077005LaFeAsO}%
  \BibitemOpen
  \bibfield  {author} {\bibinfo {author} {\bibfnamefont {H.-H.}\ \bibnamefont
  {Klauss}}, \bibinfo {author} {\bibfnamefont {H.}~\bibnamefont {Luetkens}},
  \bibinfo {author} {\bibfnamefont {R.}~\bibnamefont {Klingeler}}, \bibinfo
  {author} {\bibfnamefont {C.}~\bibnamefont {Hess}}, \bibinfo {author}
  {\bibfnamefont {F.~J.}\ \bibnamefont {Litterst}}, \bibinfo {author}
  {\bibfnamefont {M.}~\bibnamefont {Kraken}}, \bibinfo {author} {\bibfnamefont
  {M.~M.}\ \bibnamefont {Korshunov}}, \bibinfo {author} {\bibfnamefont
  {I.}~\bibnamefont {Eremin}}, \bibinfo {author} {\bibfnamefont {S.-L.}\
  \bibnamefont {Drechsler}}, \bibinfo {author} {\bibfnamefont {R.}~\bibnamefont
  {Khasanov}}, \bibinfo {author} {\bibfnamefont {A.}~\bibnamefont {Amato}},
  \bibinfo {author} {\bibfnamefont {J.}~\bibnamefont {Hamann-Borrero}},
  \bibinfo {author} {\bibfnamefont {N.}~\bibnamefont {Leps}}, \bibinfo {author}
  {\bibfnamefont {A.}~\bibnamefont {Kondrat}}, \bibinfo {author} {\bibfnamefont
  {G.}~\bibnamefont {Behr}}, \bibinfo {author} {\bibfnamefont {J.}~\bibnamefont
  {Werner}}, \ and\ \bibinfo {author} {\bibfnamefont {B.}~\bibnamefont
  {B\"uchner}},\ }\href {\doibase 10.1103/PhysRevLett.101.077005} {\bibfield
  {journal} {\bibinfo  {journal} {Phys. Rev. Lett.}\ }\textbf {\bibinfo
  {volume} {101}},\ \bibinfo {pages} {077005} (\bibinfo {year}
  {2008})}\BibitemShut {NoStop}%
\bibitem [{\citenamefont {McGuire}\ \emph {et~al.}(2008)\citenamefont
  {McGuire}, \citenamefont {Christianson}, \citenamefont {Sefat}, \citenamefont
  {Sales}, \citenamefont {Lumsden}, \citenamefont {Jin}, \citenamefont
  {Payzant}, \citenamefont {Mandrus}, \citenamefont {Luan}, \citenamefont
  {Keppens}, \citenamefont {Varadarajan}, \citenamefont {Brill}, \citenamefont
  {Hermann}, \citenamefont {Sougrati}, \citenamefont {Grandjean},\ and\
  \citenamefont {Long}}]{McGuirePRB.78.094517LaFeAsO}%
  \BibitemOpen
  \bibfield  {author} {\bibinfo {author} {\bibfnamefont {M.~A.}\ \bibnamefont
  {McGuire}}, \bibinfo {author} {\bibfnamefont {A.~D.}\ \bibnamefont
  {Christianson}}, \bibinfo {author} {\bibfnamefont {A.~S.}\ \bibnamefont
  {Sefat}}, \bibinfo {author} {\bibfnamefont {B.~C.}\ \bibnamefont {Sales}},
  \bibinfo {author} {\bibfnamefont {M.~D.}\ \bibnamefont {Lumsden}}, \bibinfo
  {author} {\bibfnamefont {R.}~\bibnamefont {Jin}}, \bibinfo {author}
  {\bibfnamefont {E.~A.}\ \bibnamefont {Payzant}}, \bibinfo {author}
  {\bibfnamefont {D.}~\bibnamefont {Mandrus}}, \bibinfo {author} {\bibfnamefont
  {Y.}~\bibnamefont {Luan}}, \bibinfo {author} {\bibfnamefont {V.}~\bibnamefont
  {Keppens}}, \bibinfo {author} {\bibfnamefont {V.}~\bibnamefont
  {Varadarajan}}, \bibinfo {author} {\bibfnamefont {J.~W.}\ \bibnamefont
  {Brill}}, \bibinfo {author} {\bibfnamefont {R.~P.}\ \bibnamefont {Hermann}},
  \bibinfo {author} {\bibfnamefont {M.~T.}\ \bibnamefont {Sougrati}}, \bibinfo
  {author} {\bibfnamefont {F.}~\bibnamefont {Grandjean}}, \ and\ \bibinfo
  {author} {\bibfnamefont {G.~J.}\ \bibnamefont {Long}},\ }\href {\doibase
  10.1103/PhysRevB.78.094517} {\bibfield  {journal} {\bibinfo  {journal} {Phys.
  Rev. B}\ }\textbf {\bibinfo {volume} {78}},\ \bibinfo {pages} {094517}
  (\bibinfo {year} {2008})}\BibitemShut {NoStop}%
\bibitem [{\citenamefont {Kitao}\ \emph {et~al.}(2008)\citenamefont {Kitao},
  \citenamefont {Kobayashi}, \citenamefont {Higashitaniguchi}, \citenamefont
  {Saito}, \citenamefont {Kamihara}, \citenamefont {Hirano}, \citenamefont
  {Mitsui}, \citenamefont {Hosono},\ and\ \citenamefont
  {Seto}}]{KitaoJPSJ2008}%
  \BibitemOpen
  \bibfield  {author} {\bibinfo {author} {\bibfnamefont {S.}~\bibnamefont
  {Kitao}}, \bibinfo {author} {\bibfnamefont {Y.}~\bibnamefont {Kobayashi}},
  \bibinfo {author} {\bibfnamefont {S.}~\bibnamefont {Higashitaniguchi}},
  \bibinfo {author} {\bibfnamefont {M.}~\bibnamefont {Saito}}, \bibinfo
  {author} {\bibfnamefont {Y.}~\bibnamefont {Kamihara}}, \bibinfo {author}
  {\bibfnamefont {M.}~\bibnamefont {Hirano}}, \bibinfo {author} {\bibfnamefont
  {T.}~\bibnamefont {Mitsui}}, \bibinfo {author} {\bibfnamefont
  {H.}~\bibnamefont {Hosono}}, \ and\ \bibinfo {author} {\bibfnamefont
  {M.}~\bibnamefont {Seto}},\ }\href {\doibase 10.1143/JPSJ.77.103706}
  {\bibfield  {journal} {\bibinfo  {journal} {Journal of the Physical Society
  of Japan}\ }\textbf {\bibinfo {volume} {77}},\ \bibinfo {pages} {103706}
  (\bibinfo {year} {2008})}\BibitemShut {NoStop}%
\bibitem [{\citenamefont {Mizuguchi}\ \emph {et~al.}(2010)\citenamefont
  {Mizuguchi}, \citenamefont {Furubayashi}, \citenamefont {Deguchi},
  \citenamefont {Tsuda}, \citenamefont {Yamaguchi},\ and\ \citenamefont
  {Takano}}]{Mizuguchi2010}%
  \BibitemOpen
  \bibfield  {author} {\bibinfo {author} {\bibfnamefont {Y.}~\bibnamefont
  {Mizuguchi}}, \bibinfo {author} {\bibfnamefont {T.}~\bibnamefont
  {Furubayashi}}, \bibinfo {author} {\bibfnamefont {K.}~\bibnamefont
  {Deguchi}}, \bibinfo {author} {\bibfnamefont {S.}~\bibnamefont {Tsuda}},
  \bibinfo {author} {\bibfnamefont {T.}~\bibnamefont {Yamaguchi}}, \ and\
  \bibinfo {author} {\bibfnamefont {Y.}~\bibnamefont {Takano}},\ }\href
  {\doibase https://doi.org/10.1016/j.physc.2009.11.142} {\bibfield  {journal}
  {\bibinfo  {journal} {Physica C: Superconductivity and its Applications}\
  }\textbf {\bibinfo {volume} {470}},\ \bibinfo {pages} {S338} (\bibinfo {year}
  {2010})},\ \bibinfo {note} {proceedings of the 9th International Conference
  on Materials and Mechanisms of Superconductivity}\BibitemShut {NoStop}%
\bibitem [{\citenamefont {Hansen}\ \emph {et~al.}(2023)\citenamefont {Hansen},
  \citenamefont {Vaney}, \citenamefont {{De Rango}}, \citenamefont {Salaün},
  \citenamefont {Tencé}, \citenamefont {Nassif},\ and\ \citenamefont
  {Toulemonde}}]{HANSEN2023169281}%
  \BibitemOpen
  \bibfield  {author} {\bibinfo {author} {\bibfnamefont {M.}~\bibnamefont
  {Hansen}}, \bibinfo {author} {\bibfnamefont {J.-B.}\ \bibnamefont {Vaney}},
  \bibinfo {author} {\bibfnamefont {P.}~\bibnamefont {{De Rango}}}, \bibinfo
  {author} {\bibfnamefont {M.}~\bibnamefont {Salaün}}, \bibinfo {author}
  {\bibfnamefont {S.}~\bibnamefont {Tencé}}, \bibinfo {author} {\bibfnamefont
  {V.}~\bibnamefont {Nassif}}, \ and\ \bibinfo {author} {\bibfnamefont
  {P.}~\bibnamefont {Toulemonde}},\ }\href {\doibase
  https://doi.org/10.1016/j.jallcom.2023.169281} {\bibfield  {journal}
  {\bibinfo  {journal} {Journal of Alloys and Compounds}\ }\textbf {\bibinfo
  {volume} {945}},\ \bibinfo {pages} {169281} (\bibinfo {year}
  {2023})}\BibitemShut {NoStop}%
\bibitem [{\citenamefont {Kamusella}\ \emph {et~al.}(2017)\citenamefont
  {Kamusella}, \citenamefont {To~Lai}, \citenamefont {Harnagea}, \citenamefont
  {Beck}, \citenamefont {Pachmayr}, \citenamefont {Singh~Thakur},\ and\
  \citenamefont {Klauss}}]{RevMossbauerPnictides}%
  \BibitemOpen
  \bibfield  {author} {\bibinfo {author} {\bibfnamefont {S.}~\bibnamefont
  {Kamusella}}, \bibinfo {author} {\bibfnamefont {K.}~\bibnamefont {To~Lai}},
  \bibinfo {author} {\bibfnamefont {L.}~\bibnamefont {Harnagea}}, \bibinfo
  {author} {\bibfnamefont {R.}~\bibnamefont {Beck}}, \bibinfo {author}
  {\bibfnamefont {U.}~\bibnamefont {Pachmayr}}, \bibinfo {author}
  {\bibfnamefont {G.}~\bibnamefont {Singh~Thakur}}, \ and\ \bibinfo {author}
  {\bibfnamefont {H.-H.}\ \bibnamefont {Klauss}},\ }\href {\doibase
  10.1002/pssb.201600160} {\bibfield  {journal} {\bibinfo  {journal} {physica
  status solidi (b)}\ }\textbf {\bibinfo {volume} {254}},\ \bibinfo {pages}
  {1600160} (\bibinfo {year} {2017})}\BibitemShut {NoStop}%
\bibitem [{\citenamefont {Fultz}(2011)}]{FlutzRevMoss}%
  \BibitemOpen
  \bibfield  {author} {\bibinfo {author} {\bibfnamefont {B.}~\bibnamefont
  {Fultz}},\ }\enquote {\bibinfo {title} {Characterization of materials},}\ \
  (\bibinfo  {publisher} {John Wiley, New York},\ \bibinfo {year} {2011})\
  Chap.\ \bibinfo {chapter} {M\"ossbauer Spectrometry}, pp.\ \bibinfo {pages}
  {816--834}\BibitemShut {NoStop}%
\bibitem [{\citenamefont {L\"ubbers}\ \emph {et~al.}(1999)\citenamefont
  {L\"ubbers}, \citenamefont {Wortmann},\ and\ \citenamefont
  {Gr\"unsteudel}}]{LubbersHI1999}%
  \BibitemOpen
  \bibfield  {author} {\bibinfo {author} {\bibfnamefont {R.}~\bibnamefont
  {L\"ubbers}}, \bibinfo {author} {\bibfnamefont {G.}~\bibnamefont {Wortmann}},
  \ and\ \bibinfo {author} {\bibfnamefont {H.~F.}\ \bibnamefont
  {Gr\"unsteudel}},\ }\href {https://doi.org/10.1023/A:1017032125551}
  {\bibfield  {journal} {\bibinfo  {journal} {Hyperfine Interactions}\ }\textbf
  {\bibinfo {volume} {123}},\ \bibinfo {pages} {529} (\bibinfo {year}
  {1999})}\BibitemShut {NoStop}%
\bibitem [{\citenamefont {Adler}\ \emph {et~al.}(2019)\citenamefont {Adler},
  \citenamefont {Medvedev}, \citenamefont {Naumov}, \citenamefont {Mohitkar},
  \citenamefont {R\"uffer}, \citenamefont {Jansen},\ and\ \citenamefont
  {Felser}}]{PhysRevB.99.134443}%
  \BibitemOpen
  \bibfield  {author} {\bibinfo {author} {\bibfnamefont {P.}~\bibnamefont
  {Adler}}, \bibinfo {author} {\bibfnamefont {S.~A.}\ \bibnamefont {Medvedev}},
  \bibinfo {author} {\bibfnamefont {P.~G.}\ \bibnamefont {Naumov}}, \bibinfo
  {author} {\bibfnamefont {S.}~\bibnamefont {Mohitkar}}, \bibinfo {author}
  {\bibfnamefont {R.}~\bibnamefont {R\"uffer}}, \bibinfo {author}
  {\bibfnamefont {M.}~\bibnamefont {Jansen}}, \ and\ \bibinfo {author}
  {\bibfnamefont {C.}~\bibnamefont {Felser}},\ }\href {\doibase
  10.1103/PhysRevB.99.134443} {\bibfield  {journal} {\bibinfo  {journal} {Phys.
  Rev. B}\ }\textbf {\bibinfo {volume} {99}},\ \bibinfo {pages} {134443}
  (\bibinfo {year} {2019})}\BibitemShut {NoStop}%
\bibitem [{\citenamefont {Adler}\ \emph {et~al.}(2020)\citenamefont {Adler},
  \citenamefont {Medvedev}, \citenamefont {Valldor}, \citenamefont {Naumov},
  \citenamefont {ElGhazali},\ and\ \citenamefont
  {R\"uffer}}]{PhysRevB.101.094433}%
  \BibitemOpen
  \bibfield  {author} {\bibinfo {author} {\bibfnamefont {P.}~\bibnamefont
  {Adler}}, \bibinfo {author} {\bibfnamefont {S.~A.}\ \bibnamefont {Medvedev}},
  \bibinfo {author} {\bibfnamefont {M.}~\bibnamefont {Valldor}}, \bibinfo
  {author} {\bibfnamefont {P.~G.}\ \bibnamefont {Naumov}}, \bibinfo {author}
  {\bibfnamefont {M.~A.}\ \bibnamefont {ElGhazali}}, \ and\ \bibinfo {author}
  {\bibfnamefont {R.}~\bibnamefont {R\"uffer}},\ }\href {\doibase
  10.1103/PhysRevB.101.094433} {\bibfield  {journal} {\bibinfo  {journal}
  {Phys. Rev. B}\ }\textbf {\bibinfo {volume} {101}},\ \bibinfo {pages}
  {094433} (\bibinfo {year} {2020})}\BibitemShut {NoStop}%
\bibitem [{\citenamefont {Adler}\ \emph {et~al.}(2022)\citenamefont {Adler},
  \citenamefont {Reehuis}, \citenamefont {St\"u\ss{}er}, \citenamefont
  {Medvedev}, \citenamefont {Nicklas}, \citenamefont {Peets}, \citenamefont
  {Bertinshaw}, \citenamefont {Christensen}, \citenamefont {Etter},
  \citenamefont {Hoser}, \citenamefont {Schr\"oder}, \citenamefont {Merz},
  \citenamefont {Schnelle}, \citenamefont {Schulz}, \citenamefont {Mu},
  \citenamefont {Bessas}, \citenamefont {Chumakov}, \citenamefont {Jansen},\
  and\ \citenamefont {Felser}}]{PhysRevB.105.054417}%
  \BibitemOpen
  \bibfield  {author} {\bibinfo {author} {\bibfnamefont {P.}~\bibnamefont
  {Adler}}, \bibinfo {author} {\bibfnamefont {M.}~\bibnamefont {Reehuis}},
  \bibinfo {author} {\bibfnamefont {N.}~\bibnamefont {St\"u\ss{}er}}, \bibinfo
  {author} {\bibfnamefont {S.~A.}\ \bibnamefont {Medvedev}}, \bibinfo {author}
  {\bibfnamefont {M.}~\bibnamefont {Nicklas}}, \bibinfo {author} {\bibfnamefont
  {D.~C.}\ \bibnamefont {Peets}}, \bibinfo {author} {\bibfnamefont
  {J.}~\bibnamefont {Bertinshaw}}, \bibinfo {author} {\bibfnamefont {C.~K.}\
  \bibnamefont {Christensen}}, \bibinfo {author} {\bibfnamefont
  {M.}~\bibnamefont {Etter}}, \bibinfo {author} {\bibfnamefont
  {A.}~\bibnamefont {Hoser}}, \bibinfo {author} {\bibfnamefont
  {L.}~\bibnamefont {Schr\"oder}}, \bibinfo {author} {\bibfnamefont
  {P.}~\bibnamefont {Merz}}, \bibinfo {author} {\bibfnamefont {W.}~\bibnamefont
  {Schnelle}}, \bibinfo {author} {\bibfnamefont {A.}~\bibnamefont {Schulz}},
  \bibinfo {author} {\bibfnamefont {Q.}~\bibnamefont {Mu}}, \bibinfo {author}
  {\bibfnamefont {D.}~\bibnamefont {Bessas}}, \bibinfo {author} {\bibfnamefont
  {A.}~\bibnamefont {Chumakov}}, \bibinfo {author} {\bibfnamefont
  {M.}~\bibnamefont {Jansen}}, \ and\ \bibinfo {author} {\bibfnamefont
  {C.}~\bibnamefont {Felser}},\ }\href {\doibase 10.1103/PhysRevB.105.054417}
  {\bibfield  {journal} {\bibinfo  {journal} {Phys. Rev. B}\ }\textbf {\bibinfo
  {volume} {105}},\ \bibinfo {pages} {054417} (\bibinfo {year}
  {2022})}\BibitemShut {NoStop}%
\bibitem [{\citenamefont {Potapkin}\ \emph {et~al.}(2012)\citenamefont
  {Potapkin}, \citenamefont {Chumakov}, \citenamefont {Smirnov}, \citenamefont
  {Celse}, \citenamefont {R{\"{u}}ffer}, \citenamefont {McCammon},\ and\
  \citenamefont {Dubrovinsky}}]{Potapkin:vv5038}%
  \BibitemOpen
  \bibfield  {author} {\bibinfo {author} {\bibfnamefont {V.}~\bibnamefont
  {Potapkin}}, \bibinfo {author} {\bibfnamefont {A.~I.}\ \bibnamefont
  {Chumakov}}, \bibinfo {author} {\bibfnamefont {G.~V.}\ \bibnamefont
  {Smirnov}}, \bibinfo {author} {\bibfnamefont {J.-P.}\ \bibnamefont {Celse}},
  \bibinfo {author} {\bibfnamefont {R.}~\bibnamefont {R{\"{u}}ffer}}, \bibinfo
  {author} {\bibfnamefont {C.}~\bibnamefont {McCammon}}, \ and\ \bibinfo
  {author} {\bibfnamefont {L.}~\bibnamefont {Dubrovinsky}},\ }\href {\doibase
  10.1107/S0909049512015579} {\bibfield  {journal} {\bibinfo  {journal}
  {Journal of Synchrotron Radiation}\ }\textbf {\bibinfo {volume} {19}},\
  \bibinfo {pages} {559} (\bibinfo {year} {2012})}\BibitemShut {NoStop}%
\bibitem [{\citenamefont {Smirnov}\ \emph {et~al.}(1997)\citenamefont
  {Smirnov}, \citenamefont {van B\"urck}, \citenamefont {Chumakov},
  \citenamefont {Baron},\ and\ \citenamefont {R\"uffer}}]{PhysRevB.55.5811}%
  \BibitemOpen
  \bibfield  {author} {\bibinfo {author} {\bibfnamefont {G.~V.}\ \bibnamefont
  {Smirnov}}, \bibinfo {author} {\bibfnamefont {U.}~\bibnamefont {van
  B\"urck}}, \bibinfo {author} {\bibfnamefont {A.~I.}\ \bibnamefont
  {Chumakov}}, \bibinfo {author} {\bibfnamefont {A.~Q.~R.}\ \bibnamefont
  {Baron}}, \ and\ \bibinfo {author} {\bibfnamefont {R.}~\bibnamefont
  {R\"uffer}},\ }\href {\doibase 10.1103/PhysRevB.55.5811} {\bibfield
  {journal} {\bibinfo  {journal} {Phys. Rev. B}\ }\textbf {\bibinfo {volume}
  {55}},\ \bibinfo {pages} {5811} (\bibinfo {year} {1997})}\BibitemShut
  {NoStop}%
\bibitem [{\citenamefont {R{\"{u}}ffer}\ and\ \citenamefont
  {Chumakov}(1996)}]{ID18ref}%
  \BibitemOpen
  \bibfield  {author} {\bibinfo {author} {\bibfnamefont {R.}~\bibnamefont
  {R{\"{u}}ffer}}\ and\ \bibinfo {author} {\bibfnamefont {A.~I.}\ \bibnamefont
  {Chumakov}},\ }\href {\doibase 10.1007/BF02150199} {\bibfield  {journal}
  {\bibinfo  {journal} {Hyperfine Interactions}\ }\textbf {\bibinfo {volume}
  {97-98}},\ \bibinfo {pages} {589} (\bibinfo {year} {1996})}\BibitemShut
  {NoStop}%
\bibitem [{\citenamefont {Prescher}\ \emph {et~al.}(2012)\citenamefont
  {Prescher}, \citenamefont {McCammon},\ and\ \citenamefont
  {Dubrovinsky}}]{Prescher:he5543}%
  \BibitemOpen
  \bibfield  {author} {\bibinfo {author} {\bibfnamefont {C.}~\bibnamefont
  {Prescher}}, \bibinfo {author} {\bibfnamefont {C.}~\bibnamefont {McCammon}},
  \ and\ \bibinfo {author} {\bibfnamefont {L.}~\bibnamefont {Dubrovinsky}},\
  }\href {\doibase 10.1107/S0021889812004979} {\bibfield  {journal} {\bibinfo
  {journal} {Journal of Applied Crystallography}\ }\textbf {\bibinfo {volume}
  {45}},\ \bibinfo {pages} {329} (\bibinfo {year} {2012})}\BibitemShut
  {NoStop}%
\bibitem [{\citenamefont {Kawakami}\ \emph {et~al.}(2009)\citenamefont
  {Kawakami}, \citenamefont {Kamatani}, \citenamefont {Okada}, \citenamefont
  {Takahashi}, \citenamefont {Nasu}, \citenamefont {Kamihara}, \citenamefont
  {Hirano},\ and\ \citenamefont {Hosono}}]{Kawakami2009_jpsj}%
  \BibitemOpen
  \bibfield  {author} {\bibinfo {author} {\bibfnamefont {T.}~\bibnamefont
  {Kawakami}}, \bibinfo {author} {\bibfnamefont {T.}~\bibnamefont {Kamatani}},
  \bibinfo {author} {\bibfnamefont {H.}~\bibnamefont {Okada}}, \bibinfo
  {author} {\bibfnamefont {H.}~\bibnamefont {Takahashi}}, \bibinfo {author}
  {\bibfnamefont {S.}~\bibnamefont {Nasu}}, \bibinfo {author} {\bibfnamefont
  {Y.}~\bibnamefont {Kamihara}}, \bibinfo {author} {\bibfnamefont
  {M.}~\bibnamefont {Hirano}}, \ and\ \bibinfo {author} {\bibfnamefont
  {H.}~\bibnamefont {Hosono}},\ }\href {\doibase 10.1143/JPSJ.78.123703}
  {\bibfield  {journal} {\bibinfo  {journal} {Journal of the Physical Society
  of Japan}\ }\textbf {\bibinfo {volume} {78}},\ \bibinfo {pages} {123703}
  (\bibinfo {year} {2009})}\BibitemShut {NoStop}%
\bibitem [{\citenamefont {Lebert}\ \emph {et~al.}(2019)\citenamefont {Lebert},
  \citenamefont {Gorni}, \citenamefont {Casula}, \citenamefont {Klotz},
  \citenamefont {Baudelet}, \citenamefont {Ablett}, \citenamefont {Hansen},
  \citenamefont {Juhin}, \citenamefont {Polian}, \citenamefont {Munsch},
  \citenamefont {Le~Marchand}, \citenamefont {Zhang}, \citenamefont {Rueff},\
  and\ \citenamefont {d{\textquoteright}Astuto}}]{Lebert20280}%
  \BibitemOpen
  \bibfield  {author} {\bibinfo {author} {\bibfnamefont {B.~W.}\ \bibnamefont
  {Lebert}}, \bibinfo {author} {\bibfnamefont {T.}~\bibnamefont {Gorni}},
  \bibinfo {author} {\bibfnamefont {M.}~\bibnamefont {Casula}}, \bibinfo
  {author} {\bibfnamefont {S.}~\bibnamefont {Klotz}}, \bibinfo {author}
  {\bibfnamefont {F.}~\bibnamefont {Baudelet}}, \bibinfo {author}
  {\bibfnamefont {J.~M.}\ \bibnamefont {Ablett}}, \bibinfo {author}
  {\bibfnamefont {T.~C.}\ \bibnamefont {Hansen}}, \bibinfo {author}
  {\bibfnamefont {A.}~\bibnamefont {Juhin}}, \bibinfo {author} {\bibfnamefont
  {A.}~\bibnamefont {Polian}}, \bibinfo {author} {\bibfnamefont
  {P.}~\bibnamefont {Munsch}}, \bibinfo {author} {\bibfnamefont
  {G.}~\bibnamefont {Le~Marchand}}, \bibinfo {author} {\bibfnamefont
  {Z.}~\bibnamefont {Zhang}}, \bibinfo {author} {\bibfnamefont {J.-P.}\
  \bibnamefont {Rueff}}, \ and\ \bibinfo {author} {\bibfnamefont
  {M.}~\bibnamefont {d{\textquoteright}Astuto}},\ }\href {\doibase
  10.1073/pnas.1904575116} {\bibfield  {journal} {\bibinfo  {journal}
  {Proceedings of the National Academy of Sciences}\ }\textbf {\bibinfo
  {volume} {116}},\ \bibinfo {pages} {20280} (\bibinfo {year}
  {2019})}\BibitemShut {NoStop}%
\bibitem [{\citenamefont {Lebert}\ \emph {et~al.}(2018)\citenamefont {Lebert},
  \citenamefont {Bal\'edent}, \citenamefont {Toulemonde}, \citenamefont
  {Ablett},\ and\ \citenamefont {Rueff}}]{Lebert_FeSe}%
  \BibitemOpen
  \bibfield  {author} {\bibinfo {author} {\bibfnamefont {B.~W.}\ \bibnamefont
  {Lebert}}, \bibinfo {author} {\bibfnamefont {V.}~\bibnamefont {Bal\'edent}},
  \bibinfo {author} {\bibfnamefont {P.}~\bibnamefont {Toulemonde}}, \bibinfo
  {author} {\bibfnamefont {J.~M.}\ \bibnamefont {Ablett}}, \ and\ \bibinfo
  {author} {\bibfnamefont {J.-P.}\ \bibnamefont {Rueff}},\ }\href {\doibase
  10.1103/PhysRevB.97.180503} {\bibfield  {journal} {\bibinfo  {journal} {Phys.
  Rev. B}\ }\textbf {\bibinfo {volume} {97}},\ \bibinfo {pages} {180503}
  (\bibinfo {year} {2018})}\BibitemShut {NoStop}%
\bibitem [{\citenamefont {Bessas}\ \emph {et~al.}(2020)\citenamefont {Bessas},
  \citenamefont {Sergueev}, \citenamefont {Glazyrin}, \citenamefont {Strohm},
  \citenamefont {Kupenko}, \citenamefont {Merkel}, \citenamefont {Long},
  \citenamefont {Grandjean}, \citenamefont {Chumakov},\ and\ \citenamefont
  {R\"uffer}}]{PhysRevB.101.035112}%
  \BibitemOpen
  \bibfield  {author} {\bibinfo {author} {\bibfnamefont {D.}~\bibnamefont
  {Bessas}}, \bibinfo {author} {\bibfnamefont {I.}~\bibnamefont {Sergueev}},
  \bibinfo {author} {\bibfnamefont {K.}~\bibnamefont {Glazyrin}}, \bibinfo
  {author} {\bibfnamefont {C.}~\bibnamefont {Strohm}}, \bibinfo {author}
  {\bibfnamefont {I.}~\bibnamefont {Kupenko}}, \bibinfo {author} {\bibfnamefont
  {D.~G.}\ \bibnamefont {Merkel}}, \bibinfo {author} {\bibfnamefont {G.~J.}\
  \bibnamefont {Long}}, \bibinfo {author} {\bibfnamefont {F.}~\bibnamefont
  {Grandjean}}, \bibinfo {author} {\bibfnamefont {A.~I.}\ \bibnamefont
  {Chumakov}}, \ and\ \bibinfo {author} {\bibfnamefont {R.}~\bibnamefont
  {R\"uffer}},\ }\href {\doibase 10.1103/PhysRevB.101.035112} {\bibfield
  {journal} {\bibinfo  {journal} {Phys. Rev. B}\ }\textbf {\bibinfo {volume}
  {101}},\ \bibinfo {pages} {035112} (\bibinfo {year} {2020})}\BibitemShut
  {NoStop}%
\bibitem [{\citenamefont {de~la Cruz}\ \emph {et~al.}(2008)\citenamefont {de~la
  Cruz}, \citenamefont {Huang}, \citenamefont {Lynn}, \citenamefont {Li},
  \citenamefont {Ratcliff~II}, \citenamefont {Zarestky}, \citenamefont {Mook},
  \citenamefont {Chen}, \citenamefont {Luo}, \citenamefont {Wang},\ and\
  \citenamefont {Dai}}]{delaCruz2008}%
  \BibitemOpen
  \bibfield  {author} {\bibinfo {author} {\bibfnamefont {C.}~\bibnamefont
  {de~la Cruz}}, \bibinfo {author} {\bibfnamefont {Q.}~\bibnamefont {Huang}},
  \bibinfo {author} {\bibfnamefont {J.~W.}\ \bibnamefont {Lynn}}, \bibinfo
  {author} {\bibfnamefont {J.}~\bibnamefont {Li}}, \bibinfo {author}
  {\bibfnamefont {W.}~\bibnamefont {Ratcliff~II}}, \bibinfo {author}
  {\bibfnamefont {J.~L.}\ \bibnamefont {Zarestky}}, \bibinfo {author}
  {\bibfnamefont {H.~A.}\ \bibnamefont {Mook}}, \bibinfo {author}
  {\bibfnamefont {G.~F.}\ \bibnamefont {Chen}}, \bibinfo {author}
  {\bibfnamefont {J.~L.}\ \bibnamefont {Luo}}, \bibinfo {author} {\bibfnamefont
  {N.~L.}\ \bibnamefont {Wang}}, \ and\ \bibinfo {author} {\bibfnamefont
  {P.}~\bibnamefont {Dai}},\ }\href {\doibase 10.1038/nature07057} {\bibfield
  {journal} {\bibinfo  {journal} {Nature}\ }\textbf {\bibinfo {volume} {453}},\
  \bibinfo {pages} {899} (\bibinfo {year} {2008})}\BibitemShut {NoStop}%
\bibitem [{\citenamefont {Layek}\ \emph {et~al.}(2023)\citenamefont {Layek},
  \citenamefont {Hansen}, \citenamefont {Vaney}, \citenamefont {Toulemonde},
  \citenamefont {Tencé}, \citenamefont {Boullay}, \citenamefont {Cano},\ and\
  \citenamefont {Méasson}}]{layek2023lattice}%
  \BibitemOpen
  \bibfield  {author} {\bibinfo {author} {\bibfnamefont {S.}~\bibnamefont
  {Layek}}, \bibinfo {author} {\bibfnamefont {M.~F.}\ \bibnamefont {Hansen}},
  \bibinfo {author} {\bibfnamefont {J.-B.}\ \bibnamefont {Vaney}}, \bibinfo
  {author} {\bibfnamefont {P.}~\bibnamefont {Toulemonde}}, \bibinfo {author}
  {\bibfnamefont {S.}~\bibnamefont {Tencé}}, \bibinfo {author} {\bibfnamefont
  {P.}~\bibnamefont {Boullay}}, \bibinfo {author} {\bibfnamefont
  {A.}~\bibnamefont {Cano}}, \ and\ \bibinfo {author} {\bibfnamefont {M.-A.}\
  \bibnamefont {Méasson}},\ }\href@noop {} {} (\bibinfo {year} {2023}),\
  \Eprint {http://arxiv.org/abs/2307.12610} {arXiv:2307.12610
  [cond-mat.supr-con]} \BibitemShut {NoStop}%
\bibitem [{\citenamefont {Qureshi}\ \emph {et~al.}(2010)\citenamefont
  {Qureshi}, \citenamefont {Drees}, \citenamefont {Werner}, \citenamefont
  {Wurmehl}, \citenamefont {Hess}, \citenamefont {Klingeler}, \citenamefont
  {B\"uchner}, \citenamefont {Fern\'andez-D\'{\i}az},\ and\ \citenamefont
  {Braden}}]{Qureshi_PRB_2010}%
  \BibitemOpen
  \bibfield  {author} {\bibinfo {author} {\bibfnamefont {N.}~\bibnamefont
  {Qureshi}}, \bibinfo {author} {\bibfnamefont {Y.}~\bibnamefont {Drees}},
  \bibinfo {author} {\bibfnamefont {J.}~\bibnamefont {Werner}}, \bibinfo
  {author} {\bibfnamefont {S.}~\bibnamefont {Wurmehl}}, \bibinfo {author}
  {\bibfnamefont {C.}~\bibnamefont {Hess}}, \bibinfo {author} {\bibfnamefont
  {R.}~\bibnamefont {Klingeler}}, \bibinfo {author} {\bibfnamefont
  {B.}~\bibnamefont {B\"uchner}}, \bibinfo {author} {\bibfnamefont {M.~T.}\
  \bibnamefont {Fern\'andez-D\'{\i}az}}, \ and\ \bibinfo {author}
  {\bibfnamefont {M.}~\bibnamefont {Braden}},\ }\href {\doibase
  10.1103/PhysRevB.82.184521} {\bibfield  {journal} {\bibinfo  {journal} {Phys.
  Rev. B}\ }\textbf {\bibinfo {volume} {82}},\ \bibinfo {pages} {184521}
  (\bibinfo {year} {2010})}\BibitemShut {NoStop}%
\bibitem [{\citenamefont {Tabuchi}\ \emph {et~al.}(2010)\citenamefont
  {Tabuchi}, \citenamefont {Li}, \citenamefont {Oka}, \citenamefont {Chen},
  \citenamefont {Kawasaki}, \citenamefont {Luo}, \citenamefont {Wang},\ and\
  \citenamefont {Zheng}}]{Tabuchi}%
  \BibitemOpen
  \bibfield  {author} {\bibinfo {author} {\bibfnamefont {T.}~\bibnamefont
  {Tabuchi}}, \bibinfo {author} {\bibfnamefont {Z.}~\bibnamefont {Li}},
  \bibinfo {author} {\bibfnamefont {T.}~\bibnamefont {Oka}}, \bibinfo {author}
  {\bibfnamefont {G.~F.}\ \bibnamefont {Chen}}, \bibinfo {author}
  {\bibfnamefont {S.}~\bibnamefont {Kawasaki}}, \bibinfo {author}
  {\bibfnamefont {J.~L.}\ \bibnamefont {Luo}}, \bibinfo {author} {\bibfnamefont
  {N.~L.}\ \bibnamefont {Wang}}, \ and\ \bibinfo {author} {\bibfnamefont
  {G.-q.}\ \bibnamefont {Zheng}},\ }\href {\doibase 10.1103/PhysRevB.81.140509}
  {\bibfield  {journal} {\bibinfo  {journal} {Phys. Rev. B}\ }\textbf {\bibinfo
  {volume} {81}},\ \bibinfo {pages} {140509} (\bibinfo {year}
  {2010})}\BibitemShut {NoStop}%
\bibitem [{\citenamefont {Zhou}\ \emph {et~al.}(2020)\citenamefont {Zhou},
  \citenamefont {Scherer}, \citenamefont {Mayaffre}, \citenamefont
  {Toulemonde}, \citenamefont {Ma}, \citenamefont {Li}, \citenamefont
  {Andersen},\ and\ \citenamefont {Julien}}]{Zhou2020}%
  \BibitemOpen
  \bibfield  {author} {\bibinfo {author} {\bibfnamefont {R.}~\bibnamefont
  {Zhou}}, \bibinfo {author} {\bibfnamefont {D.~D.}\ \bibnamefont {Scherer}},
  \bibinfo {author} {\bibfnamefont {H.}~\bibnamefont {Mayaffre}}, \bibinfo
  {author} {\bibfnamefont {P.}~\bibnamefont {Toulemonde}}, \bibinfo {author}
  {\bibfnamefont {M.}~\bibnamefont {Ma}}, \bibinfo {author} {\bibfnamefont
  {Y.}~\bibnamefont {Li}}, \bibinfo {author} {\bibfnamefont {B.~M.}\
  \bibnamefont {Andersen}}, \ and\ \bibinfo {author} {\bibfnamefont {M.-H.}\
  \bibnamefont {Julien}},\ }\href {\doibase 10.1038/s41535-020-00295-1}
  {\bibfield  {journal} {\bibinfo  {journal} {npj Quantum Materials}\ }\textbf
  {\bibinfo {volume} {5}},\ \bibinfo {pages} {93} (\bibinfo {year}
  {2020})}\BibitemShut {NoStop}%
\bibitem [{\citenamefont {Wiecki}\ \emph {et~al.}(2021)\citenamefont {Wiecki},
  \citenamefont {Zhou}, \citenamefont {Julien}, \citenamefont {B\"ohmer},\ and\
  \citenamefont {Schmalian}}]{Wiecki2021}%
  \BibitemOpen
  \bibfield  {author} {\bibinfo {author} {\bibfnamefont {P.}~\bibnamefont
  {Wiecki}}, \bibinfo {author} {\bibfnamefont {R.}~\bibnamefont {Zhou}},
  \bibinfo {author} {\bibfnamefont {M.-H.}\ \bibnamefont {Julien}}, \bibinfo
  {author} {\bibfnamefont {A.~E.}\ \bibnamefont {B\"ohmer}}, \ and\ \bibinfo
  {author} {\bibfnamefont {J.}~\bibnamefont {Schmalian}},\ }\href {\doibase
  10.1103/PhysRevB.104.125134} {\bibfield  {journal} {\bibinfo  {journal}
  {Phys. Rev. B}\ }\textbf {\bibinfo {volume} {104}},\ \bibinfo {pages}
  {125134} (\bibinfo {year} {2021})}\BibitemShut {NoStop}%
\bibitem [{\citenamefont {Yang}\ \emph {et~al.}(2015)\citenamefont {Yang},
  \citenamefont {Zhou}, \citenamefont {Wei}, \citenamefont {Yang},
  \citenamefont {LI}, \citenamefont {Zhao},\ and\ \citenamefont
  {Zheng}}]{Yang}%
  \BibitemOpen
  \bibfield  {author} {\bibinfo {author} {\bibfnamefont {J.}~\bibnamefont
  {Yang}}, \bibinfo {author} {\bibfnamefont {R.}~\bibnamefont {Zhou}}, \bibinfo
  {author} {\bibfnamefont {L.-L.}\ \bibnamefont {Wei}}, \bibinfo {author}
  {\bibfnamefont {H.-X.}\ \bibnamefont {Yang}}, \bibinfo {author}
  {\bibfnamefont {J.-Q.}\ \bibnamefont {LI}}, \bibinfo {author} {\bibfnamefont
  {Z.-X.}\ \bibnamefont {Zhao}}, \ and\ \bibinfo {author} {\bibfnamefont
  {G.-Q.}\ \bibnamefont {Zheng}},\ }\href {\doibase
  10.1088/0256-307X/32/10/107401} {\bibfield  {journal} {\bibinfo  {journal}
  {Chinese Physics Letters}\ }\textbf {\bibinfo {volume} {32}},\ \bibinfo {eid}
  {107401} (\bibinfo {year} {2015})}\BibitemShut {NoStop}%
\bibitem [{\citenamefont {Ma}\ \emph {et~al.}(2011)\citenamefont {Ma},
  \citenamefont {Ji}, \citenamefont {Dai}, \citenamefont {He}, \citenamefont
  {Wang}, \citenamefont {Chen}, \citenamefont {Normand},\ and\ \citenamefont
  {Yu}}]{Ma11}%
  \BibitemOpen
  \bibfield  {author} {\bibinfo {author} {\bibfnamefont {L.}~\bibnamefont
  {Ma}}, \bibinfo {author} {\bibfnamefont {G.~F.}\ \bibnamefont {Ji}}, \bibinfo
  {author} {\bibfnamefont {J.}~\bibnamefont {Dai}}, \bibinfo {author}
  {\bibfnamefont {J.~B.}\ \bibnamefont {He}}, \bibinfo {author} {\bibfnamefont
  {D.~M.}\ \bibnamefont {Wang}}, \bibinfo {author} {\bibfnamefont {G.~F.}\
  \bibnamefont {Chen}}, \bibinfo {author} {\bibfnamefont {B.}~\bibnamefont
  {Normand}}, \ and\ \bibinfo {author} {\bibfnamefont {W.}~\bibnamefont {Yu}},\
  }\href {\doibase 10.1103/PhysRevB.84.220505} {\bibfield  {journal} {\bibinfo
  {journal} {Phys. Rev. B}\ }\textbf {\bibinfo {volume} {84}},\ \bibinfo
  {pages} {220505} (\bibinfo {year} {2011})}\BibitemShut {NoStop}%
\bibitem [{\citenamefont {Ning}\ \emph {et~al.}(2010)\citenamefont {Ning},
  \citenamefont {Ahilan}, \citenamefont {Imai}, \citenamefont {Sefat},
  \citenamefont {McGuire}, \citenamefont {Sales}, \citenamefont {Mandrus},
  \citenamefont {Cheng}, \citenamefont {Shen},\ and\ \citenamefont
  {Wen}}]{Ning}%
  \BibitemOpen
  \bibfield  {author} {\bibinfo {author} {\bibfnamefont {F.~L.}\ \bibnamefont
  {Ning}}, \bibinfo {author} {\bibfnamefont {K.}~\bibnamefont {Ahilan}},
  \bibinfo {author} {\bibfnamefont {T.}~\bibnamefont {Imai}}, \bibinfo {author}
  {\bibfnamefont {A.~S.}\ \bibnamefont {Sefat}}, \bibinfo {author}
  {\bibfnamefont {M.~A.}\ \bibnamefont {McGuire}}, \bibinfo {author}
  {\bibfnamefont {B.~C.}\ \bibnamefont {Sales}}, \bibinfo {author}
  {\bibfnamefont {D.}~\bibnamefont {Mandrus}}, \bibinfo {author} {\bibfnamefont
  {P.}~\bibnamefont {Cheng}}, \bibinfo {author} {\bibfnamefont
  {B.}~\bibnamefont {Shen}}, \ and\ \bibinfo {author} {\bibfnamefont {H.-H.}\
  \bibnamefont {Wen}},\ }\href {\doibase 10.1103/PhysRevLett.104.037001}
  {\bibfield  {journal} {\bibinfo  {journal} {Phys. Rev. Lett.}\ }\textbf
  {\bibinfo {volume} {104}},\ \bibinfo {pages} {037001} (\bibinfo {year}
  {2010})}\BibitemShut {NoStop}%
\bibitem [{\citenamefont {Ma}\ and\ \citenamefont {Yu}(2013)}]{Ma}%
  \BibitemOpen
  \bibfield  {author} {\bibinfo {author} {\bibfnamefont {L.}~\bibnamefont
  {Ma}}\ and\ \bibinfo {author} {\bibfnamefont {W.-Q.}\ \bibnamefont {Yu}},\
  }\href {\doibase 10.1088/1674-1056/22/8/087414} {\bibfield  {journal}
  {\bibinfo  {journal} {Chinese Physics B}\ }\textbf {\bibinfo {volume} {22}},\
  \bibinfo {pages} {087414} (\bibinfo {year} {2013})}\BibitemShut {NoStop}%
\bibitem [{\citenamefont {Carretta}\ and\ \citenamefont
  {Prando}(2020)}]{Carretta}%
  \BibitemOpen
  \bibfield  {author} {\bibinfo {author} {\bibfnamefont {P.}~\bibnamefont
  {Carretta}}\ and\ \bibinfo {author} {\bibfnamefont {G.}~\bibnamefont
  {Prando}},\ }\href {\doibase 10.1007/s40766-019-0001-1} {\bibfield  {journal}
  {\bibinfo  {journal} {La Rivista del Nuovo Cimento}\ }\textbf {\bibinfo
  {volume} {43}},\ \bibinfo {pages} {1} (\bibinfo {year} {2020})}\BibitemShut
  {NoStop}%
\bibitem [{\citenamefont {Iye}\ \emph {et~al.}(2012)\citenamefont {Iye},
  \citenamefont {Nakai}, \citenamefont {Kitagawa}, \citenamefont {Ishida},
  \citenamefont {Kasahara}, \citenamefont {Shibauchi}, \citenamefont
  {Matsuda},\ and\ \citenamefont {Terashima}}]{Iye2012}%
  \BibitemOpen
  \bibfield  {author} {\bibinfo {author} {\bibfnamefont {T.}~\bibnamefont
  {Iye}}, \bibinfo {author} {\bibfnamefont {Y.}~\bibnamefont {Nakai}}, \bibinfo
  {author} {\bibfnamefont {S.}~\bibnamefont {Kitagawa}}, \bibinfo {author}
  {\bibfnamefont {K.}~\bibnamefont {Ishida}}, \bibinfo {author} {\bibfnamefont
  {S.}~\bibnamefont {Kasahara}}, \bibinfo {author} {\bibfnamefont
  {T.}~\bibnamefont {Shibauchi}}, \bibinfo {author} {\bibfnamefont
  {Y.}~\bibnamefont {Matsuda}}, \ and\ \bibinfo {author} {\bibfnamefont
  {T.}~\bibnamefont {Terashima}},\ }\href {\doibase 10.1103/PhysRevB.85.184505}
  {\bibfield  {journal} {\bibinfo  {journal} {Phys. Rev. B}\ }\textbf {\bibinfo
  {volume} {85}},\ \bibinfo {pages} {184505} (\bibinfo {year}
  {2012})}\BibitemShut {NoStop}%
\bibitem [{\citenamefont {Hu}\ \emph {et~al.}(2015)\citenamefont {Hu},
  \citenamefont {Lu}, \citenamefont {Zhang}, \citenamefont {Luo}, \citenamefont
  {Li}, \citenamefont {Wang}, \citenamefont {Chen}, \citenamefont {Han},
  \citenamefont {Banjara}, \citenamefont {Sapkota}, \citenamefont {Kreyssig},
  \citenamefont {Goldman}, \citenamefont {Yamani}, \citenamefont {Niedermayer},
  \citenamefont {Skoulatos}, \citenamefont {Georgii}, \citenamefont {Keller},
  \citenamefont {Wang}, \citenamefont {Yu},\ and\ \citenamefont
  {Dai}}]{Hu2015}%
  \BibitemOpen
  \bibfield  {author} {\bibinfo {author} {\bibfnamefont {D.}~\bibnamefont
  {Hu}}, \bibinfo {author} {\bibfnamefont {X.}~\bibnamefont {Lu}}, \bibinfo
  {author} {\bibfnamefont {W.}~\bibnamefont {Zhang}}, \bibinfo {author}
  {\bibfnamefont {H.}~\bibnamefont {Luo}}, \bibinfo {author} {\bibfnamefont
  {S.}~\bibnamefont {Li}}, \bibinfo {author} {\bibfnamefont {P.}~\bibnamefont
  {Wang}}, \bibinfo {author} {\bibfnamefont {G.}~\bibnamefont {Chen}}, \bibinfo
  {author} {\bibfnamefont {F.}~\bibnamefont {Han}}, \bibinfo {author}
  {\bibfnamefont {S.~R.}\ \bibnamefont {Banjara}}, \bibinfo {author}
  {\bibfnamefont {A.}~\bibnamefont {Sapkota}}, \bibinfo {author} {\bibfnamefont
  {A.}~\bibnamefont {Kreyssig}}, \bibinfo {author} {\bibfnamefont {A.~I.}\
  \bibnamefont {Goldman}}, \bibinfo {author} {\bibfnamefont {Z.}~\bibnamefont
  {Yamani}}, \bibinfo {author} {\bibfnamefont {C.}~\bibnamefont {Niedermayer}},
  \bibinfo {author} {\bibfnamefont {M.}~\bibnamefont {Skoulatos}}, \bibinfo
  {author} {\bibfnamefont {R.}~\bibnamefont {Georgii}}, \bibinfo {author}
  {\bibfnamefont {T.}~\bibnamefont {Keller}}, \bibinfo {author} {\bibfnamefont
  {P.}~\bibnamefont {Wang}}, \bibinfo {author} {\bibfnamefont {W.}~\bibnamefont
  {Yu}}, \ and\ \bibinfo {author} {\bibfnamefont {P.}~\bibnamefont {Dai}},\
  }\href {\doibase 10.1103/PhysRevLett.114.157002} {\bibfield  {journal}
  {\bibinfo  {journal} {Phys. Rev. Lett.}\ }\textbf {\bibinfo {volume} {114}},\
  \bibinfo {pages} {157002} (\bibinfo {year} {2015})}\BibitemShut {NoStop}%
\bibitem [{\citenamefont {Phejar}\ \emph {et~al.}(2016)\citenamefont {Phejar},
  \citenamefont {Paul-Boncour},\ and\ \citenamefont {Bessais}}]{PHEJAR201695}%
  \BibitemOpen
  \bibfield  {author} {\bibinfo {author} {\bibfnamefont {M.}~\bibnamefont
  {Phejar}}, \bibinfo {author} {\bibfnamefont {V.}~\bibnamefont
  {Paul-Boncour}}, \ and\ \bibinfo {author} {\bibfnamefont {L.}~\bibnamefont
  {Bessais}},\ }\href {\doibase https://doi.org/10.1016/j.jssc.2015.10.016}
  {\bibfield  {journal} {\bibinfo  {journal} {Journal of Solid State
  Chemistry}\ }\textbf {\bibinfo {volume} {233}},\ \bibinfo {pages} {95}
  (\bibinfo {year} {2016})}\BibitemShut {NoStop}%
\bibitem [{\citenamefont {Fujieda}\ \emph {et~al.}(2008)\citenamefont
  {Fujieda}, \citenamefont {Fujita}, \citenamefont {Fukamichi}, \citenamefont
  {Yamaguchi},\ and\ \citenamefont {Ohoyama}}]{Fujieda}%
  \BibitemOpen
  \bibfield  {author} {\bibinfo {author} {\bibfnamefont {S.}~\bibnamefont
  {Fujieda}}, \bibinfo {author} {\bibfnamefont {A.}~\bibnamefont {Fujita}},
  \bibinfo {author} {\bibfnamefont {K.}~\bibnamefont {Fukamichi}}, \bibinfo
  {author} {\bibfnamefont {Y.}~\bibnamefont {Yamaguchi}}, \ and\ \bibinfo
  {author} {\bibfnamefont {K.}~\bibnamefont {Ohoyama}},\ }\href {\doibase
  10.1143/JPSJ.77.074722} {\bibfield  {journal} {\bibinfo  {journal} {Journal
  of the Physical Society of Japan}\ }\textbf {\bibinfo {volume} {77}},\
  \bibinfo {pages} {074722} (\bibinfo {year} {2008})}\BibitemShut {NoStop}%
\end{thebibliography}
\end{document}